\documentclass[preprints,article,accept,moreauthors,pdftex]{Definitions/mdpi} 

\usepackage{textgreek}
\usepackage{amsmath}
\usepackage{amssymb}
\usepackage{varwidth}
\usepackage[section]{placeins}
\usepackage{subfig}
\usepackage{braket}

\firstpage{1} 
\pubvolume{1}
\issuenum{1}
\articlenumber{0}
\pubyear{2021}
\copyrightyear{2020}
\datereceived{} 
\dateaccepted{} 
\datepublished{} 
\hreflink{https://doi.org/} % If needed use \linebreak
\pdfoutput=1

\Title{Surface characterization of p-type point contact germanium detectors}

\TitleCitation{Surface characterization of p-type point contact germanium detectors}

\Author{Frank Edzards~$^{1,2}$*, Lukas Hauertmann~$^{1,2}$, Iris Abt~$^1$, Chris Gooch~$^1$, Björn Lehnert~$^3$, Xiang Liu~$^1$, \mbox{Susanne Mertens~$^{1,2}$}, David~C.~Radford~$^4$, Oliver Schulz~$^1$ and Michael Willers~$^{1,2}$}

\AuthorNames{Frank Edzards, Lukas Hauertmann, Iris Abt, Chris Gooch, Björn Lehnert, Xiang Liu, Susanne Mertens, David~C.~Radford, Oliver Schulz and Michael Willers}

\AuthorCitation{F.~Edzards \emph{et al.}}
\address{%
$^{1}$ \quad Max Planck Institute for Physics, F\"ohringer Ring 6, 80805 Munich, Germany\\
$^{2}$ \quad Technical University of Munich, Arcisstrasse 21, 80333 Munich, Germany\\
$^{3}$ \quad Lawrence Berkeley National Laboratory, 1 Cyclotron Rd, CA 94720, US\\
$^{4}$ \quad Oak Ridge National Laboratory, 1 Bethel Valley Rd, TN 37830, US}

\corres{Correspondence: edzards@mpp.mpg.de}

\abstract{P-type point contact~(PPC) germanium detectors are used in rare event and low-background searches, including neutrinoless double beta~($0\text{\textnu\textbeta\textbeta}$) decay, low-energy nuclear recoils, and coherent elastic neutrino-nucleus scattering. The detectors feature an excellent energy resolution, low detection thresholds down to the sub-keV range, and enhanced background rejection capabilities. However, due to their large passivated surface, separating the signal readout contact from the bias voltage electrode, PPC~detectors are susceptible to surface effects such as charge build-up. A~profound understanding of their response to surface events is essential. In this work, the response of a PPC~detector to alpha and beta particles hitting the passivated surface was investigated in a multi-purpose scanning test stand. It is shown that the passivated surface can accumulate charges resulting in a radial-dependent degradation of the observed event energy. In addition, it is demonstrated that the pulse shapes of surface alpha events show characteristic features which can be used to discriminate against these events.}


\begin{document}

\section{Introduction}\label{ch:introduction}
The observation of neutrinoless double beta~($0\text{\textnu\textbeta\textbeta}$) decay would have major implications on our understanding of the origin of matter in our universe. The decay violates lepton number conservation by two units, and the search for it is the most practical way to ascertain whether neutrinos are Majorana particles, i.e.~their own antiparticles~($\text{\textnu}=\bar{\text{\textnu}}$). Moreover, together with cosmological observations and direct neutrino mass measurements, it could provide information on the absolute neutrino mass scale and ordering~\cite{dolinski2019, giuliani2020}. 
\\\\
One of the most promising technologies to search for $0\text{\textnu\textbeta\textbeta}$~decay are high-purity germanium (HPGe) detectors. Germanium detectors are intrinsically pure, can be enriched to above 92\% in the double beta decaying isotope $^{76}$Ge, and provide an excellent energy resolution of about 0.1\%~FWHM~(full width at half maximum) in the region of interest around $Q_{\text{\textbeta\textbeta}}=2039\,$keV.

%#############################################################################
%#############################################################################
%#############################################################################
\section{PPC germanium detectors}\label{ch:sec1}
\begin{figure}[!h]
\centering\quad
\begin{varwidth}[b]{0.5\linewidth}
\begin{tabular}{ll}
\toprule
Property & Value \\
\midrule
Mass    				&$1.0\,$kg	\\
Inner diameter $a$		&$58.9\,$mm \\
Outer diameter $b$		&$68.9\,$mm \\
Length $c$				&$52.0\,$mm \\
Length $d$ 				&$47.0\,$mm \\
Deadlayer (Ge/Li) $e$	&$1.1\,$mm  \\
Dimple depth $f$		&$2.0\,$mm	\\
Dimple diameter $g$		&$3.2\,$mm	\\
\bottomrule
\end{tabular}
\quad
\end{varwidth}
\begin{minipage}[b]{0.5\linewidth}
\centering
\includegraphics[angle=0,width=0.75\textwidth]{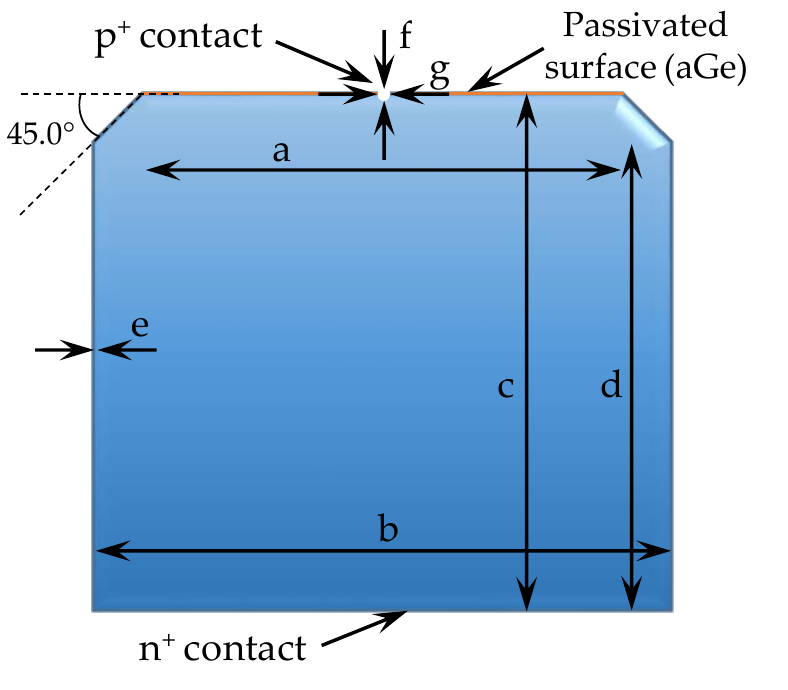}
\vspace{-2.2cm}
\end{minipage}
\caption{Sketch and parameters of the PPC~detector under study.}
\label{graph:detector_dimensions}
\end{figure}

P-type point contact~(PPC) germanium detectors are semiconductor detectors with a cylindrical shape, see \textbf{Fig.}~\ref{graph:detector_dimensions}. While the n$^+$~contact extends over the lateral and bottom detector surface, the p$^+$~readout contact has the form of a small circular well located at the center of the top surface. The size of the point contact is significantly smaller than that of traditional semi-coaxial detectors. Therefore, PPC~detectors have a lower capacitance, typically in the range of $1-2\,\text{pF}$ at full depletion, resulting in lower electronic noise and thus in a better energy resolution~\cite{luke1989,mertens2019}. Moreover, PPC~detectors can be operated at lower energy thresholds~($<1\,$keV) which makes them suitable for rare event searches at small energies~\cite{alvis2019}. Another advantage of this type of detector is the enhanced capability of applying background rejection methods based on so-called pulse shapes. This is due to the specific geometry and arrangement of the electrodes leading to a strong electric field close to the readout contact, and to a relatively low field elsewhere. As a result, the signal shape of events that deposit their energy at a single location~(single-site events like $0\text{\textnu\textbeta\textbeta}$~decay events) in the detector is almost independent of the location of the event. This can be used to discriminate these events from events where energy is deposited at multiple sites (multi-site events like Compton-scattered photons) which are a major source of background~\cite{agostini2013,edzards2020}.
\FloatBarrier
\noindent

%#############################################################################
%#############################################################################
%#############################################################################
\section{Surface events as backgrounds for $0\text{\textnu\textbeta\textbeta}$ decay searches}\label{ch:surface_backgrounds}

\subsection{Alpha backgrounds}
Events induced by alpha particles are a background for $^{76}$Ge-based $0\text{\textnu\textbeta\textbeta}$ decay searches. They are predominantly caused by the decay of radon isotopes and their progeny, particularly~$^{222}$Rn. Radon is a radioactive noble gas which is created naturally as part of the decay chains of uranium and thorium. During the production processing of a germanium detector, it is exposed to air and undergoes various mechanical and chemical treatments. A slight radon contamination of the detector~(surface) is unavoidable. Furthermore, radon contamination and outgassing of parts close to the detectors in the experimental environment can also lead to undesired alpha backgrounds.
\\\\
In the decay chain of~$^{222}$Rn, the long-lived isotope~$^{210}$Po is of major concern. During its decay to the stable isotope $^{206}$Pb, an alpha particle with an energy of~$5407.5\,$keV is emitted. If the energy of the alpha particle is degraded, it can lead to a background in the region of interest. The degradation can be caused by the alpha particle loosing energy in the material from which it is emitted, by loosing energy in layers on or close to the detector surface, or by charge trapping or charge loss due to dead regions in the detector. In germanium, the penetration depth of an alpha particle with an energy of~$E_{\alpha}\approx5\,$MeV is of the order of~$20\,\text{\textmu m}$.

\subsection{Beta backgrounds}
Beta backgrounds are particularly relevant for $0\text{\textnu\textbeta\textbeta}$~decay searches 
for which detectors are submerged in liquid argon~(LAr), such as in the Large Enriched Germanium Experiment for Neutrinoless $\text{\textbeta\textbeta}$~Decay (LEGEND)~\mbox{\cite{abgrall2017,abgrall2018,myslik2019}}. The long-lived isotope $^{42}$Ar ($T_{1/2}=32.9\,$yr) is naturally abundant in LAr when sourced from the atmosphere. It is produced by cosmogenic activation and decays via single beta decay to the short-lived daughter $^{42}$K ($T_{1/2}=12.36\,$h). The decay energy of $Q_{\text{\textbeta}}=599\,$keV is too low to create a background event in the region of interest at the $Q_{\text{\textbeta\textbeta}}$-value of $^{76}$Ge. However, subsequent beta decays of the short-lived daughter $^{42}$K with a decay energy of $Q_{\text{\textbeta}}=3525.4\,$keV are a potential source of background. 
\\\\
Within the LAr~volume, the path length of beta particles from $^{42}$K decays is less than $1.6\,$cm~\cite{agostini2020}. Hence, they are only detected if the decay happens in close proximity to the detector surface. Independent of where the beta particles hit the surface, they can lead to background events. However, when impinging on the thick n$^+$~lithium layer, surface beta events have a characteristic signal shape which can be used to discriminate against them~\cite{abgrall2020}.
\\\\
The main difference between alpha and beta particles is their penetration depth into the germanium detector. In contrast to alpha particles, electrons from beta decay penetrate deeper, typically up to several millimeters, depending on their energy. Therefore, not all beta particles show the characteristics of events close to the surface, so-called surface effects.

%#############################################################################
\subsection{Surface effects and signal development}\label{ch:surface_effects}
\begin{figure}[!h]
%\myfloatalign
\subfloat[Weighting potential in $(r,z)$~plane.] {\label{graph:galatea_wp_scatter}
\includegraphics[width=.48\linewidth]{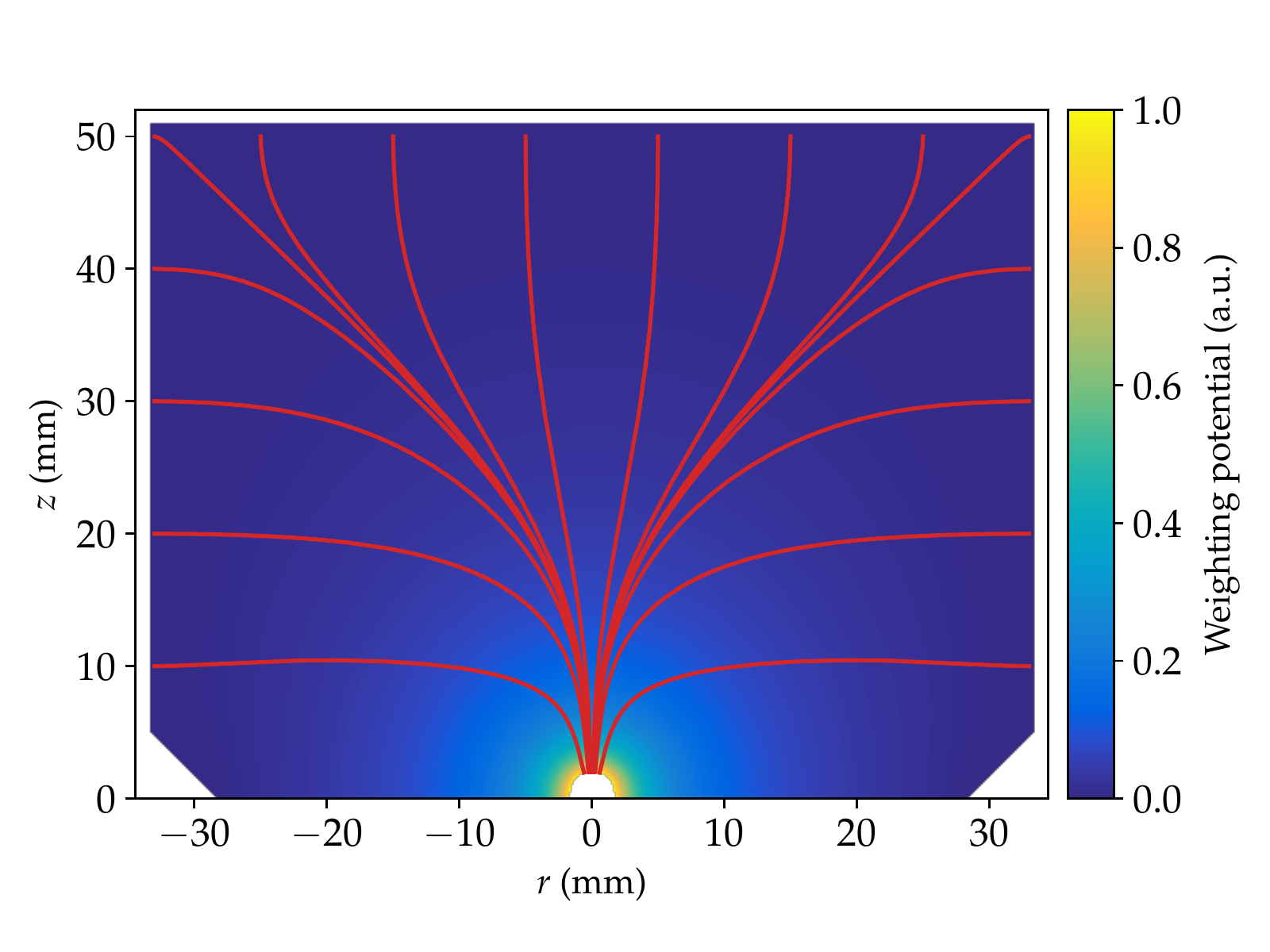}}\quad
\subfloat[Weighting potential at $z=0\,$mm.] {\label{graph:galatea_wp_z0} 
\includegraphics[width=.45\linewidth]{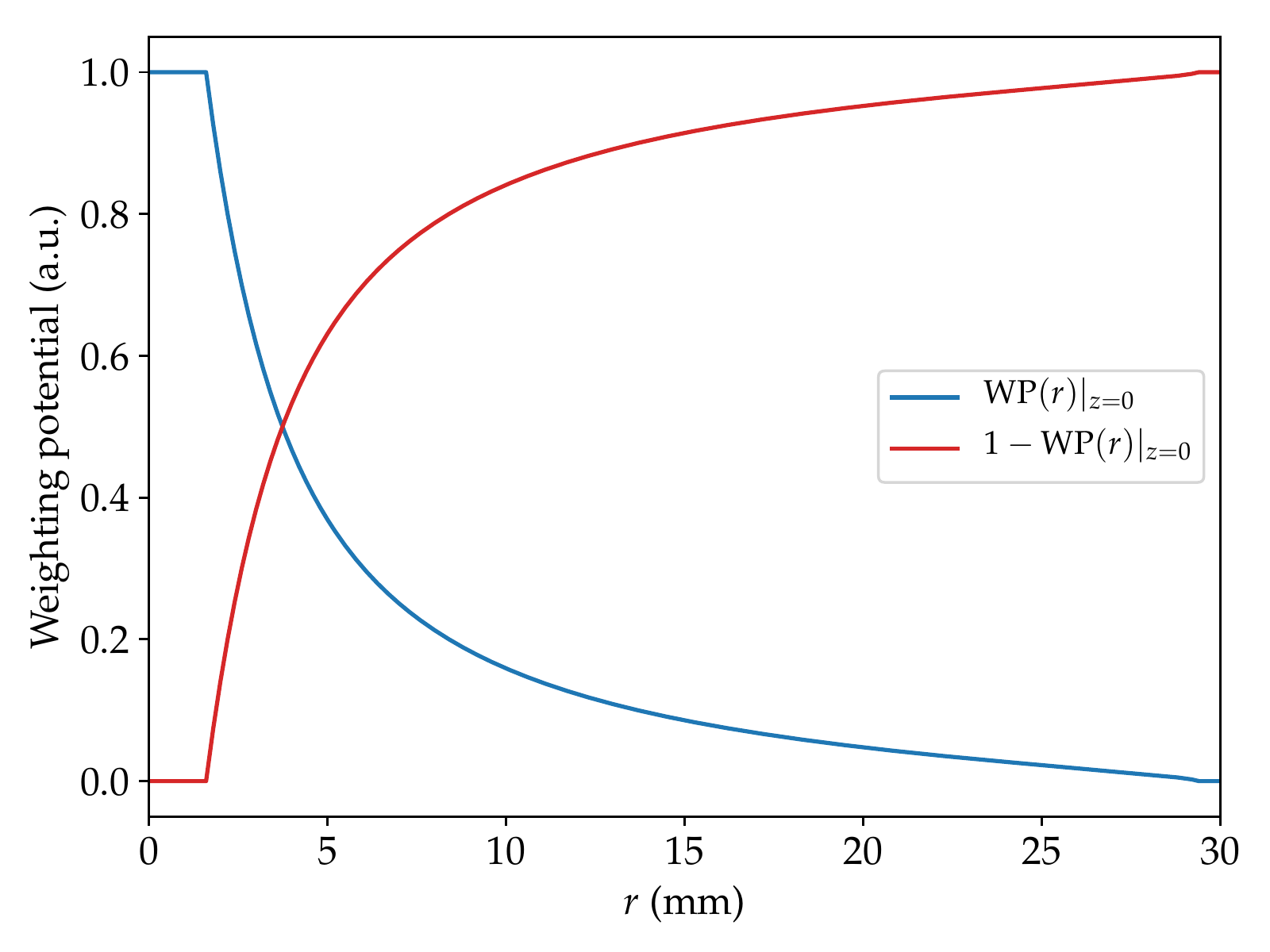}}
\caption{(a)~Weighting potential~(WP) of the investigated PPC detector in the $(r,z)$~plane. Red lines indicate the hole drift paths. (b)~Radial dependence of the WP at the passivated surface~($z=0\,$mm). The WP was simulated using \texttt{mjd\_fieldgen}~\cite{radford_siggen}.}
\label{graph:galatea_weighting_potential}
\end{figure}

Compared to most other HPGe detector geometries, PPC~detectors have a large passivated surface, usually of the order of~$30-40\,\text{cm}^2$. This surface extends over the horizontal top surface~($z=0\,$mm) excluding the p$^+$~contact, see \textbf{Fig.}~\ref{graph:detector_dimensions}. Typically, the passivated surface is made from sputtered silicon oxide or amorphous germanium~(aGe). This layer has a high resistivity and is left floating, i.e.~it is at an undefined electric potential. While the n$^+$~contact is insensitive to surface alpha events~(alpha particles cannot penetrate the few mm-thick lithium-drifted layer), beta particles entering through this surface lead to characteristically slow pulses and can be discriminated against. In contrast, the passivated surface and the point contact are highly sensitive to alpha and beta surface events.
\\\\
Since the passivation layer is left floating, it is susceptible to charge build-up. A non-zero charge on the passivated surface which for example can be induced by nearby materials at non-zero potentials, changes the electric field in the vicinity of this surface and thus affects the signal formation. Without any charge build-up, the electric field lines close to the passivated surface are almost parallel to that surface. However, in the presence of surface charges, the field has a strong perpendicular component, modifying the hole and electron drift paths.
\\\\
The signal formation of a germanium detector is described by the Shockley-Ramo theorem~\cite{shockley1938,ramo1939}. Any interaction inside the detector creates a cloud of pairs of charge carriers, i.e.~holes and electrons. These charge carriers immediately induce positive and negative mirror charges in the electrodes. The holes drift towards the p$^+$~contact, whereas the electrons drift to the n$^+$~electrode. For a deposited charge~$q$, the time-dependent signal~$S(t)$ in the p$^+$~contact is given by
\begin{align}
S(t)=q\,\left[\text{WP}(\vec{r}_{\text{h}}(t))-\text{WP}(\vec{r}_{\text{e}}(t))\right],
\label{eq:shockley_ramo}
\end{align}
where $\text{WP}(\vec{r}(t))$ denotes the weighting potential at the respective hole (electron) position~$\vec{r}_\text{h}(t)$~($\vec{r}_{\text{e}}(t)$). The weighting potential of an electrode is determined by the detector geometry and describes how strongly the charge at a given detector position couples to this electrode. For the following discussion, the p$^+$~contact is the electrode of interest. By definition, the weighting potential on this contact is one, while it is zero on the n$^+$~contact. 
\\\\
At time $t=0$, since $\vec{r}_{\text{h}}(0)=\vec{r}_{\text{e}}(0)$, the signal is $S(0)=0$. As the holes approach the p$^+$~electrode, $\text{WP}(\vec{r}_{\text{h}}(t))$ increases, see \textbf{Fig.}~\ref{graph:galatea_wp_scatter}, until the holes are collected at time~$t_\text{h}^{\text{col}}$, and $\text{WP}(\vec{r}_{\text{h}}(t))=1$ for $t\ge t_{\text{h}}^{\text{col}}$. As the electrons approach the n$^+$~electrode, $|\text{WP}(\vec{r}_{\text{e}}(t))|$ decreases until the electrons are collected at time~$t_{\text{e}}^{\text{col}}$ and $\text{WP}(\vec{r}_{\text{e}}(t))=0$. As soon as both kinds of charge carriers are collected, only the collected holes determine the signal and $S(t)=q$. 
\FloatBarrier
\noindent
The weighting potential close to the passivated detector surface strongly depends on the radius~$r$. As shown in \textbf{Fig.}~\ref{graph:galatea_wp_z0}, the weighting potential~$\text{WP}(r)|_{z=0}$ drops quickly with increasing~$r$. The term $1-\text{WP}(r)|_{z=0}$ shows the opposite behavior, i.e.~it increases with radius. 

\begin{figure}%[!h]
%\myfloatalign
\subfloat[Negative surface charges.] {\label{graph:galatea_neg_charges} 
\includegraphics[width=.5\linewidth]{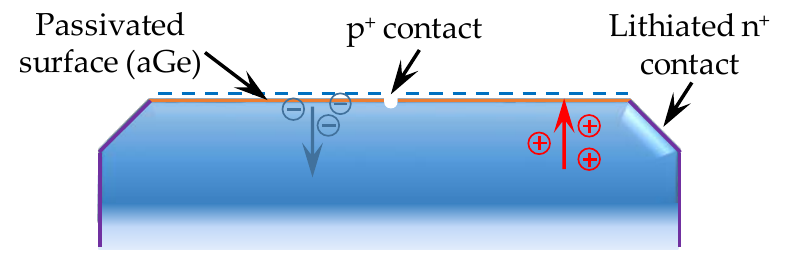}}
%\vspace{-0.3cm}
\subfloat[Positive surface charges.] {\label{graph:galatea_pos_charges} 
\includegraphics[width=.5\linewidth]{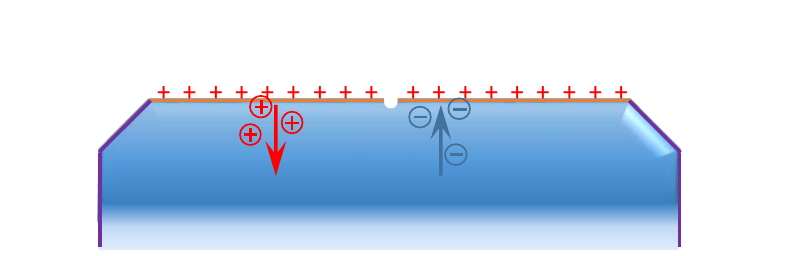}}
\caption{Effect of (a)~negative and (b)~positive charges on the passivated surface of a PPC~germanium detector. In the case of negative (positive) surface charges, holes (electrons) are attracted to the surface, whereas electrons (holes) are repelled.}
\label{graph:galatea_surface_charges}
\end{figure}

\begin{figure}[!h]
%\myfloatalign
\subfloat[Negative surface charges (\mbox{$\sigma=-0.3\cdot10^{10}\,\text{e}/\text{cm}^2$}).] {\label{graph:galatea_paths_neg_charges} 
\includegraphics[width=.48\linewidth]{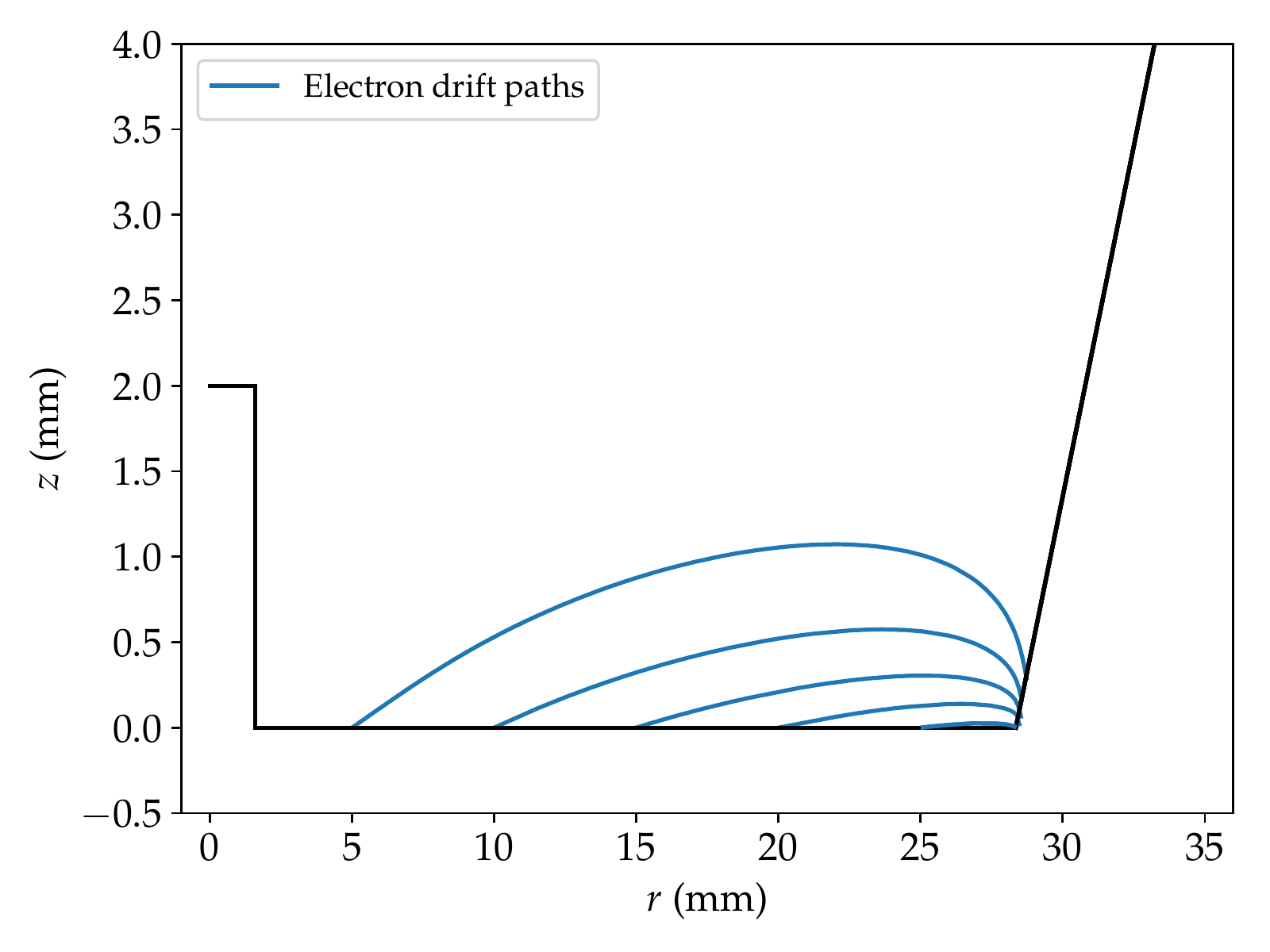}}\quad
\subfloat[Positive surface charges (\mbox{$\sigma=+0.3\cdot10^{10}\,\text{e}/\text{cm}^2$}).] {\label{graph:galatea_paths_pos_charges}
\includegraphics[width=.48\linewidth]{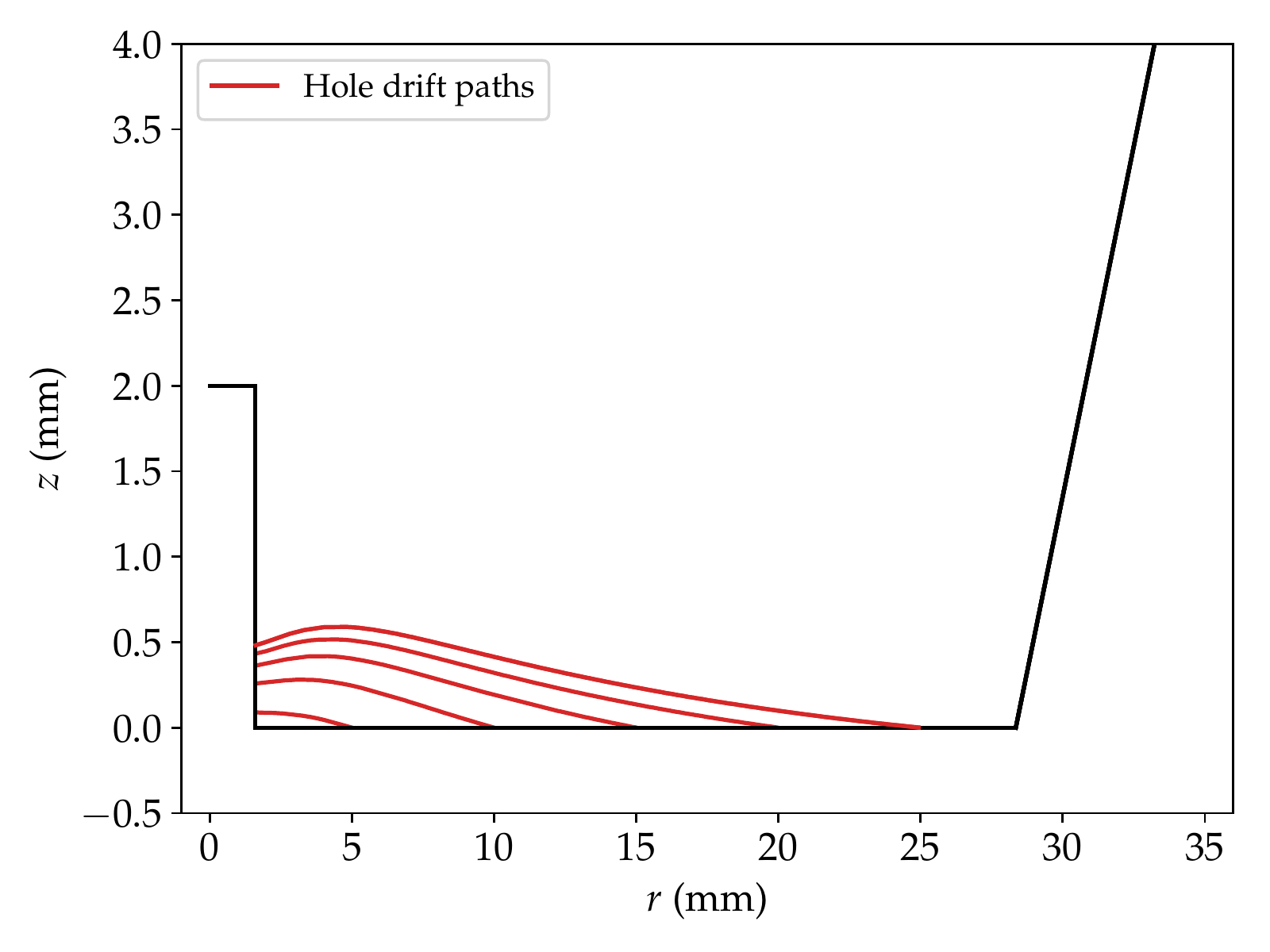}}
\caption{Simulated electron and hole drift paths for (a)~negative and (b)~positive charges on the passivated detector surface. The black line indicates the detector contour, the passivated surface is at~$z=0$. The origin of the coordinate system is at the top center of the point contact with the $z$-axis pointing towards the detector bulk. In the presence of negative (positive) surface charges, the holes (electrons) are attracted to the surface and stop drifting, while the electrons (holes) are repelled into the bulk and do not drift parallel to the surface but inside the bulk towards the respective contacts. The drift paths were simulated using \texttt{mjd\_siggen}~\cite{radford_siggen}.}
\label{graph:galatea_drift_paths_wcharges}
\end{figure}

\paragraph{\emph{Negative surface charges}}
If the passivated surface carries a negative charge, $\sigma<0$, holes which are not created in close proximity to the p$^+$~contact do not drift directly towards the p$^+$~contact, but are diverted to the surface, see \textbf{Fig.}~\ref{graph:galatea_neg_charges}. At the passivated surface, they drift very slowly parallel to this surface in the direction of the p$^+$~contact. Some holes might even get trapped and stop moving. As a result, the holes are almost stationary and are not collected, at least not within the time in which the signal is recorded. This time is tailored to normal bulk events and is too short to cover the possible collection of delayed holes. Therefore, for times~$t\ge t_{\text{e}}^{\text{col}}$, i.e.~after electron collection, the signal~$S(t)$ becomes
\begin{align}
S(t)\approx q\,\text{WP}(\vec{r}_{\text{h}}(0)).\label{eq:neg_surf_sig}
\end{align}
The larger the radius~$r$, the smaller the final signal amplitude. Due to the presence of negative surface charges, the electrons are repelled from the surface. Simulated electron drift paths are shown in \textbf{Fig.}~\ref{graph:galatea_paths_neg_charges}. The paths, which are normally almost parallel to the surface, are modified, i.e.~the electrons penetrate deeper into the bulk.
%\FloatBarrier
\noindent
%\FloatBarrier
\noindent

\paragraph{\emph{Positive surface charges}}
For a positive charge on the passivated detector surface, $\sigma>0$, the electrons created during a particle interaction are attracted to the surface, whereas the holes are repelled, see \textbf{Fig.}~\ref{graph:galatea_pos_charges}. The drift paths of the holes are modified as shown in \textbf{Fig.}~\ref{graph:galatea_paths_pos_charges}. In this case, the electrons are almost stationary and are not collected during the time in which the signal is recorded. Thus, for times~$t\ge t_{\text{h}}^\text{col}$, i.e.~after hole collection, the signal becomes
\begin{align}
S(t)\approx q\,\left[1-\text{WP}(\vec{r}_{\text{e}}(0))\right].
\end{align}
The larger the radius~$r$, the larger the final signal amplitude.

\begin{figure}[!h]
\begin{center}
%\hspace{-2cm}
\includegraphics[angle=0,width=0.73\textwidth]{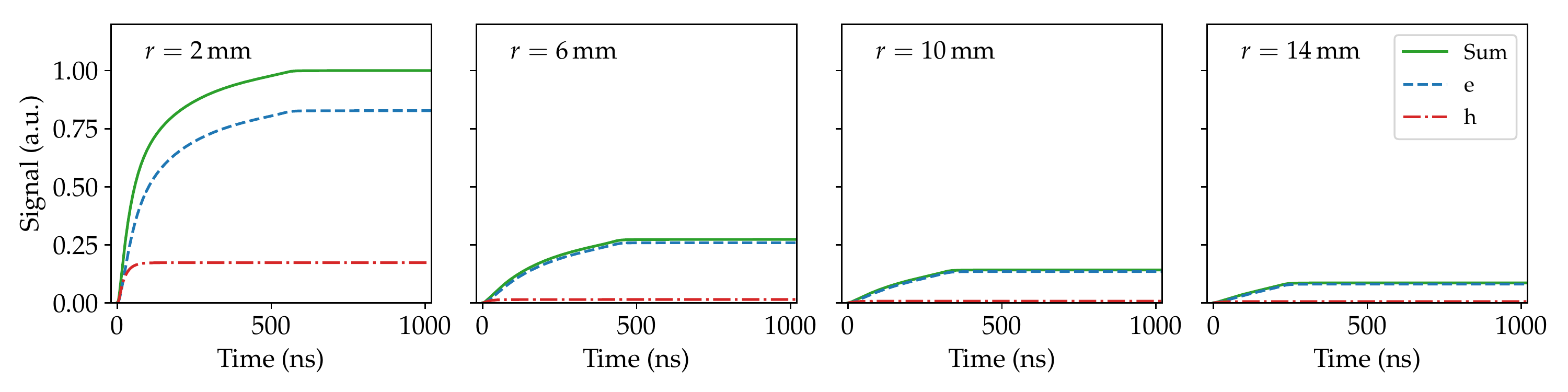}
\caption{Simulated development of the alpha event signals in the presence of negative charges on the passivated detector surface \mbox{($\sigma=-0.3\cdot10^{10}\,\text{e}/\text{cm}^2$)} for selected radial positions. Point-like charges were started at a depth of~$z_0=16\,\text{\textmu m}$. The components from holes~(h) and electrons~(e) are discussed in the text. The waveforms were simulated using \texttt{mjd\_siggen}~\cite{radford_siggen}.}
\label{graph:galatea_alpha_signals}
\end{center}
\end{figure}

\paragraph{\emph{Impact of surface charges on alpha and beta events}}
For both, negative and positive charges on the passivated detector surface, the signal amplitude is reduced. After the application of the standard calibration, the signal becomes the observed energy~$E^{\text{obs}}$. In both cases, $E^{\text{obs}}$ is smaller than the true event energy~$E^{\text{true}}$. The radial dependence of~$E^{\text{obs}}$ allows to distinguish experimentally between the two cases.
\\\\
Due to the small penetration depth of alpha particles in germanium, it is expected that basically all charge carriers are affected by the surface effects described above. Thus, in the case of negative surface charges, it is expected that~$E^{\text{obs}}$ approximately follows the radial dependence of the weighting potential. The expected signal development for a homogeneously distributed surface charge \mbox{($\sigma=-0.3\cdot10^{10}\,\text{e}/\text{cm}^2$)} at varying radii is shown in \textbf{Fig.}~\ref{graph:galatea_alpha_signals}. Point-like normalized charges created at $z_0=16\,\text{\textmu m}$ were used in the simulation. Holes created very close to the p$^+$~contact are collected quickly. However, the negative charge induced by the electrons still close to the p$^+$~contact reduces the signal amplitude. The electrons drift away from the p$^+$~contact and the effect is shown as a positive contribution. At $r=2$\,mm, the holes are fully collected and when the electrons have reached the n$^+$~contact, the signal is~$S(t)=1$. At higher radii, the holes are not collected and their signal contribution is constant from~$t=0$ on. Only the negative contribution to the signal from the electrons becomes smaller as the electrons drift. Again, this is shown as a positive contribution being almost identical to~$S(t)$ which reaches its final value of $\text{WP}(\vec{r}_\text{h}(0))$ at time~$t_\text{e}^\text{col}$, cf.~\textbf{Eq.}~(\ref{eq:neg_surf_sig}). Consequently, the observed energy~$E^{\text{obs}}$ follows the radially declining weighting potential.
%\FloatBarrier
\noindent

%#############################################################################
\subsection{Delayed charge recovery}\label{ch:dcr}
\begin{figure}[!h]
\begin{center}
\includegraphics[width=0.6\linewidth]{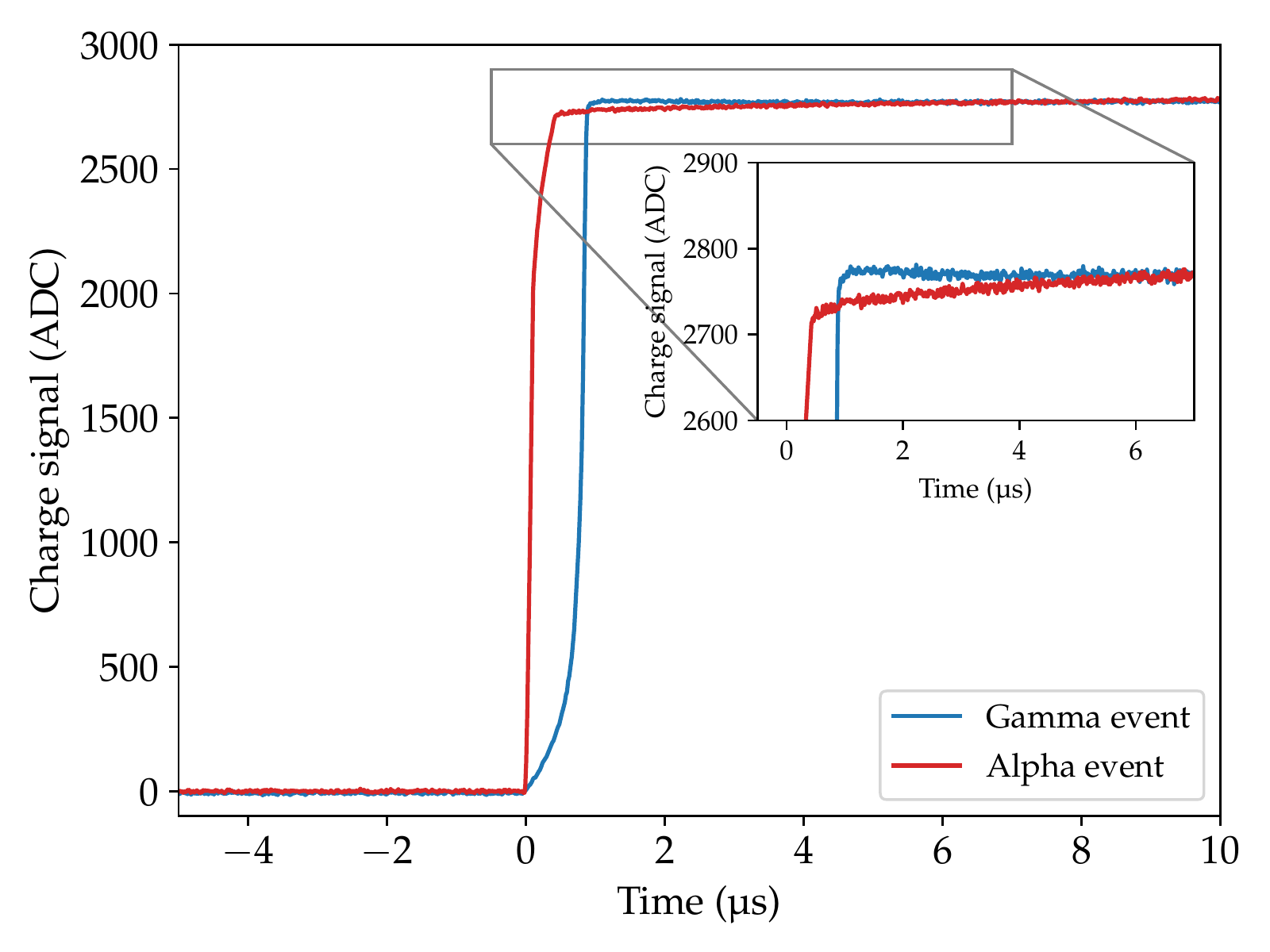}
\caption{Typical examples of a bulk gamma event~(blue curve) and of a surface alpha event~(red curve) with the same energy. The waveforms were recorded using the PPC~detector under study. The baseline and pole-zero-corrected alpha signal features a slowly rising tail~(see inset). This component can be explained as due to DCR. Due to its proximity to the signal readout electrode, the rise time of the surface alpha event is shorter (steeply rising leading edge) than the rise time of the gamma bulk event.}
\label{graph:galatea_am241_alpha_waveform_ex}
\end{center}
\end{figure}

Delayed charge recovery~(DCR) describes the phenomenon of an extra slow charge collection component for surface alpha events~\mbox{\cite{alvis2019,gruszko2017,arnquist2020}}. In the last section, the drift velocity of one kind of charge carriers was assumed to be too low to observe charge collection. DCR reflects that at least some part of the affected charge carriers are collected within the time of signal recording. Compared to events in the detector bulk~(e.g.~gamma events), the presence of a DCR component for surface alpha events modifies the tail of the signal pulse. As shown in \textbf{Fig.}~\ref{graph:galatea_am241_alpha_waveform_ex}, the tail of the pole-zero-corrected waveform\footnote{This correction addresses the decay of the signal in a charge sensitive amplifier.} still increases after the charge collection in the detector bulk can be assumed to be completed. In contrast, for a gamma event with the same energy in the detector bulk, the tail stays at a constant value. The distinct DCR~feature in the waveform makes the DCR~effect an effective tool to identify and reject surface alpha events on the passivated surface of PPC~detectors.
\FloatBarrier
\noindent
There are two mechanisms that can potentially explain the delayed collection of charges for surface alpha events:
\begin{enumerate}
\item A certain fraction of charges created during the alpha interaction is trapped in a $\mathcal{O}(\text{\textmu m})$-thick region at or near the passivated surface. In this case, the DCR effect corresponds to a slow release of these charges into the detector bulk~(with a certain release time $\tau_r$) and their subsequent drift to the electrodes.
\item Charges created on or close to the passivated surface have a significantly reduced drift velocity compared to the drift velocity in the detector bulk~\cite{cooper2012}. In this case, the DCR effect corresponds to a slow drift of charges along the passivated surface.
\end{enumerate}
Typically, the charge drift along the passivated surface takes much longer than the time, in which the waveforms are digitized. Until recently, it was assumed that the main component leading to DCR is due to the trapping and the subsequent slow release of charges at the passivated surface. In previous measurements, a charge release time on the order of several microseconds was observed. In addition, the fraction of charge released into the detector bulk was on the order of a few percent~\mbox{\cite{gruszko2017,arnquist2020}.} However, pulse shape simulations including the effects of diffusion and self-repulsion have demonstrated that surface drifts can also have an impact, cf.~\textbf{Ch.}~\ref{ch:simu_diffusion_repulsion}.
\\\\
The DCR~effect can be exploited to define a tail-based pulse shape discrimination parameter, the DCR~rate parameter. It is computed by estimating the slope~$\delta$ of the pole-zero-corrected waveform tail based on a two-point slope estimate~\mbox{\cite{alvis2019,arnquist2020}}: 
\begin{align}
\delta=\frac{y_1-y_2}{t_1-t_2}.
\end{align}
Here, $y_1, y_2$ denote average signal values, and $t_1, t_2$ average time values. The time values correspond to the average values in the intervals
\begin{align}
\text{I)}~t_{97\%}+2\,\text{\textmu s}\leq &~t\leq t_{97\%}+3\,\text{\textmu s},\\
\text{II)}~t_{\text{last}}-1\,\text{\textmu s}\leq &~t\leq t_{\text{last}},
\end{align}
where $t_{97\%}$ is the time, at which the waveform has reached $97\%$ of its maximum amplitude, and $t_{\text{last}}$ is the time corresponding to the last sample of the waveform trace. These time windows have been chosen to allow comparisons with the measurement results presented in~\mbox{\cite{alvis2019,gruszko2017,arnquist2020}.} However, it should be noted that this definition introduces a slight dependence on the trigger time in the waveform trace: The second window is not defined relative to the onset of the charge collection, but rather comprises a fixed window (last microsecond of the waveform trace).

%#############################################################################
%#############################################################################
%#############################################################################
\section{Measurements}\label{ch:measurements}

\subsection{Experimental setup}
\begin{figure}[!h]
\begin{center}
%\hspace{-2cm}
\includegraphics[angle=0,width=0.38\textwidth]{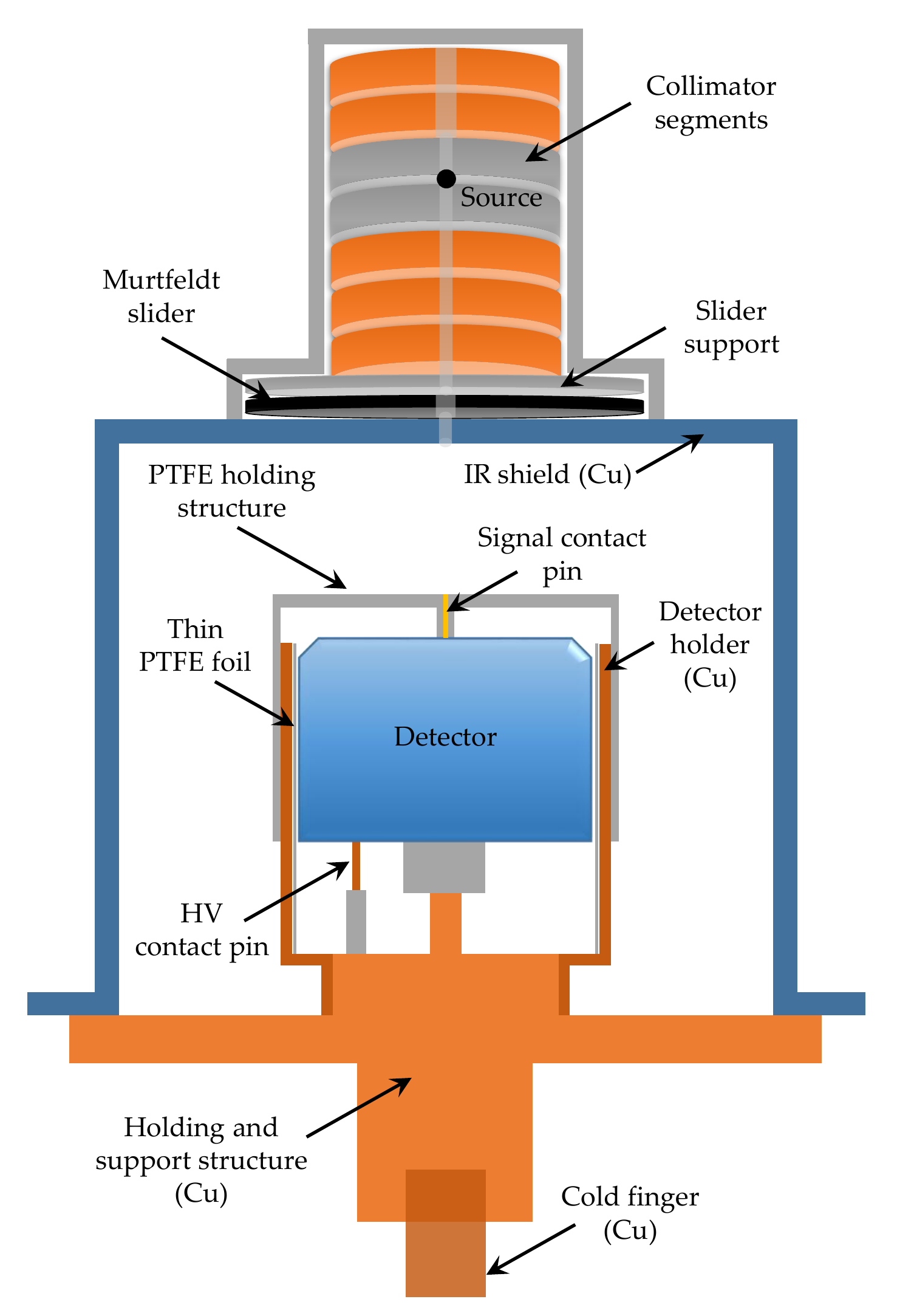}
\caption{Simplified sectional view of the PPC~detector installed in the \textsc{Galatea} test facility. The detector is mounted in a copper structure that is cooled via liquid nitrogen. It is surrounded by a copper IR~shield. For the scan measurements presented in this manuscript, the detector top surface was irradiated by a radioactive source installed in the top collimator above the IR~shield. For reasons of visual clarity, the side collimator of \textsc{Galatea}, not relevant for this work, is not shown.}
\label{graph:galatea_ponama1}
\end{center}
\end{figure}

The PPC~detector surface characterization measurements presented in this work were carried out in the \textsc{Galatea}~(GermAnium LAser TEst Apparatus) facility, a fully automated multi-purpose scanning test stand that was built to investigate bulk and surface effects of HPGe detectors~\cite{abt2015,irlbeck2014}. Due to its versatility, it allows for almost complete scans of the detector surface with alpha, beta, and gamma radiation. 
\\\\
The facility is a large customized vacuum cryostat housing the scanning stages, the germanium detector, the radioactive source(s), and the signal readout electronics. The detector under investigation can be mounted in an aluminum or a copper holding structure which is also used for its cooling. The detector is shielded against infrared~(IR) radiation by a cylindrical copper hat. The IR~shield has two slits (one on the side of the hat, one on top), along which the collimators with the radioactive sources are guided during the measurements. A system consisting of three independent stages allows an almost complete scan of the detector surface. One stage can rotate the IR~shield up to~$360^{\circ}$ with respect to the detector, facilitating azimuthal scans. The additional two linear stages are used to move the top collimator across the top surface for top scans, and the side collimator vertically for side scans.
\\\\
The detector used for this work, see \textbf{Fig.}~\ref{graph:detector_dimensions}, is a PPC~germanium detector with natural isotopic composition and properties that closely resemble those of the detectors previously operated in the \mbox{\textsc{Majorana Demonstrator}} \cite{alvis2019,aalseth2018}. To allow for a scan of the passivated surface, the detector was installed with the point contact facing up in a customized detector mount, see \textbf{Fig.}~\ref{graph:galatea_ponama1}. The n$^+$~electrode was connected to the high voltage module via a spring-loaded pin located at the detector bottom. Likewise, connection to the p$^+$~contact was established with a pogo pin which was attached to a narrow PTFE~bar mounted on top of the detector. In addition, the PTFE~holding structure was used to guide the signal cable.
\\\\
Data were acquired with a Struck $14$-bit SIS$3316$ flash ADC~(FADC) digitizing the analog signals  from the charge sensitive amplifier with a sampling frequency of~$250\,$MHz. For every signal, $5000$~samples corresponding to a total waveform trace length of~$20\,\text{\textmu s}$ were recorded. 
\FloatBarrier
\noindent

%#############################################################################
\subsection{Characterization of surface alpha interactions}\label{ch:alpha_characterization}

\subsubsection{Source configuration}
For the PPC~detector surface characterization with alpha particles, an $^{241}$Am source with an activity of $A_0=40\,$kBq and an expected FWHM of ${\sim}\,19\,$keV at the $5.5\,$MeV alpha peak was mounted in the top collimator of \textsc{Galatea}. Suitable cylindrical PTFE and copper segments were used to fill the collimator frame. Based on the collimator geometry and the source strength, an alpha rate of ${\sim}\,0.7~\text{counts/s}$ was expected at the detector top surface. In all measurements, the $^{241}$Am beam spot had an incidence of~$90^{\circ}$ on the detector surface. In close vicinity to the point contact, the beam spot was shadowed by the PTFE bar. Hence, it was not possible to take data in this region. An uncollimated $^{228}$Th source~($A_0=100\,$kBq) was additionally mounted on top of the IR~shield for energy calibration purposes, and to confirm pulse shape discrimination capabilities.
\\\\
Several radial scans at different azimuthal positions, as well as background and stability measurements were conducted. For the radial scans, a measurement time of~$2\,$hr at each scan point provided sufficiently high statistics. The detector was operated at a bias voltage of~$V_{\text{B}}=1050\,$V. 

\subsubsection{Results}\label{ch:am241_radial_dependence}
\begin{figure}[!h]
\begin{center}
\includegraphics[width=0.6\linewidth]{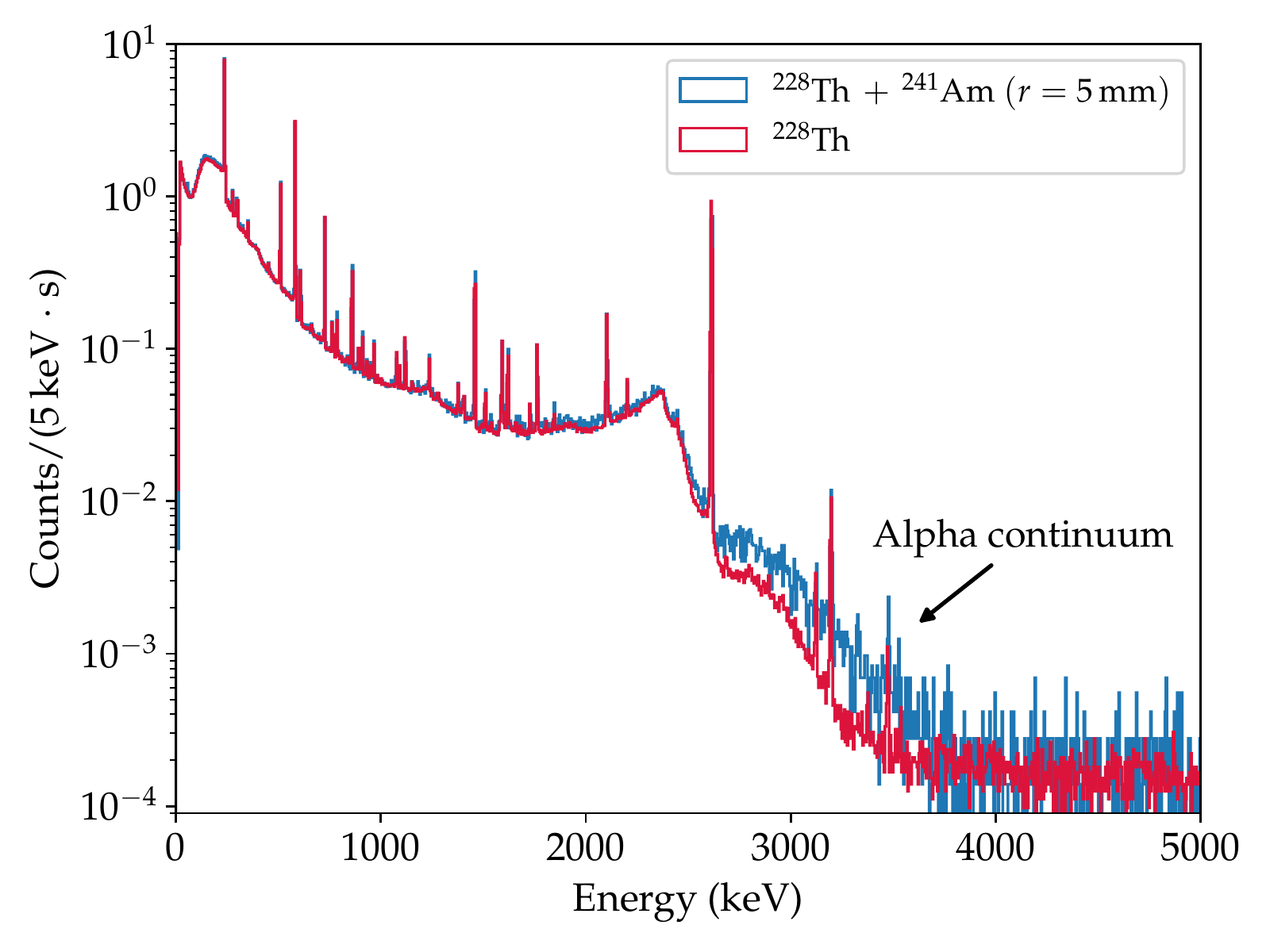}
\caption{Energy spectra as measured with the $^{228}$Th and the $^{241}$Am source at the position $r=5\,$mm (blue curve), and the $^{228}$Th source only (red curve). At higher energies, a continuum of alpha events is visible.}
\label{graph:galatea_am241_full_spectrum}
\end{center}
\end{figure}

The radial response of the PPC~detector to surface alpha events is of special interest. The energy spectra of a measurement with the $^{241}$Am source at~\mbox{$r=5\,$mm}, and of a measurement with only the $^{228}$Th source present are shown in \textbf{Fig.}~\ref{graph:galatea_am241_full_spectrum}. The contribution from the $^{228}$Th~source dominates the energy spectrum up to ${\sim}\,2.6\,$MeV. At higher energies, the measurement with the $^{241}$Am~source deviates from the $^{228}$Th-only measurement. The higher count rate is attributed to alpha events. To isolate these events from other events, radius-independent multivariate cuts were developed~\cite{edzards2021}. More specifically, cuts on various pulse shape parameters were used to exclude regions in the parameter space which did not contain alpha events. 
\FloatBarrier
\noindent

\begin{figure}[!h]
%\myfloatalign
\subfloat[Energy histograms of alpha events.] {\label{graph:galatea_am241_alpha_energy_histo}
\includegraphics[width=.48\linewidth]{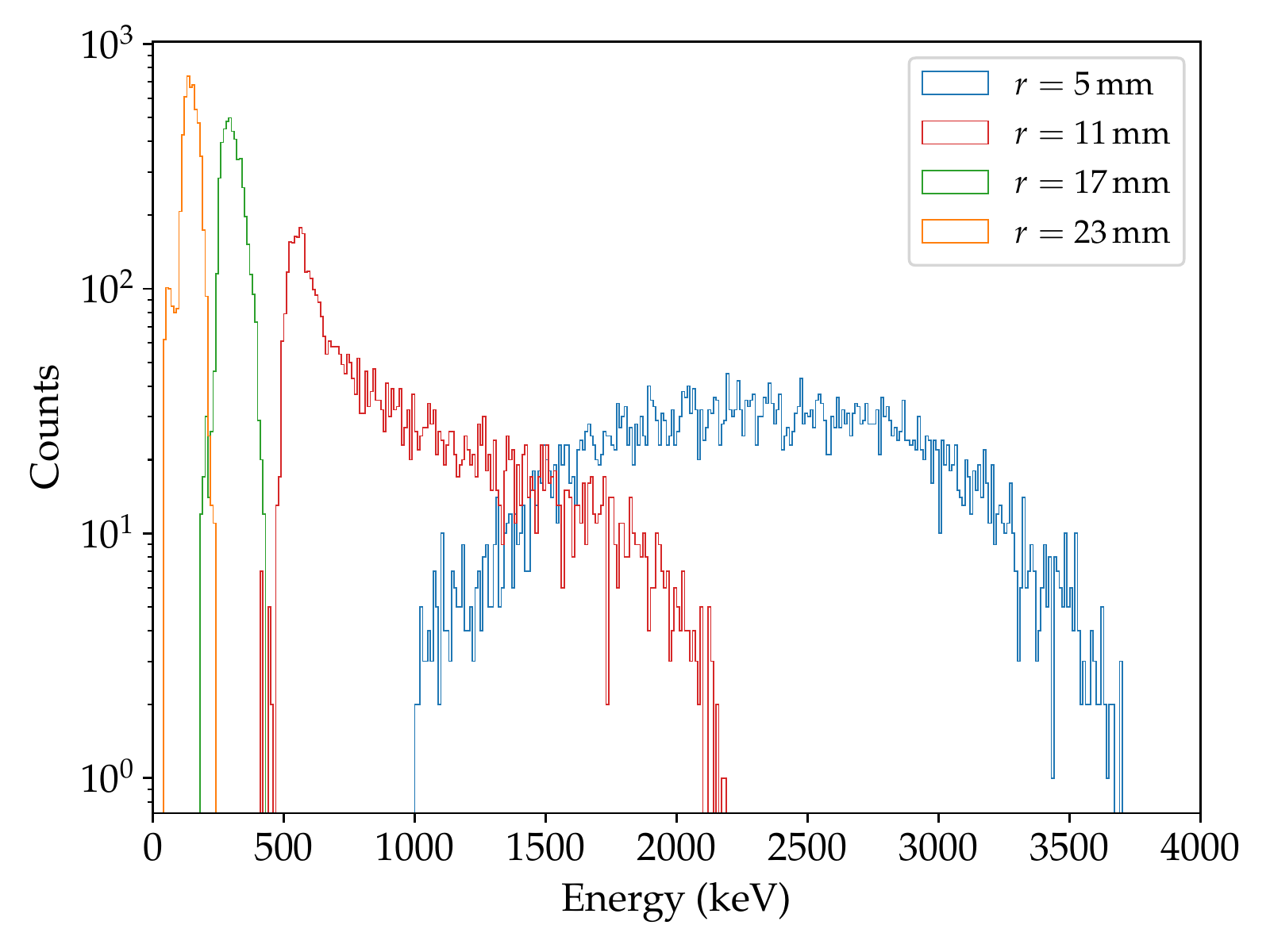}}\quad
\subfloat[Radial energy dependence of alpha energies.] {\label{graph:galatea_am241_alpha_radial_energy}
\includegraphics[width=.48\linewidth]{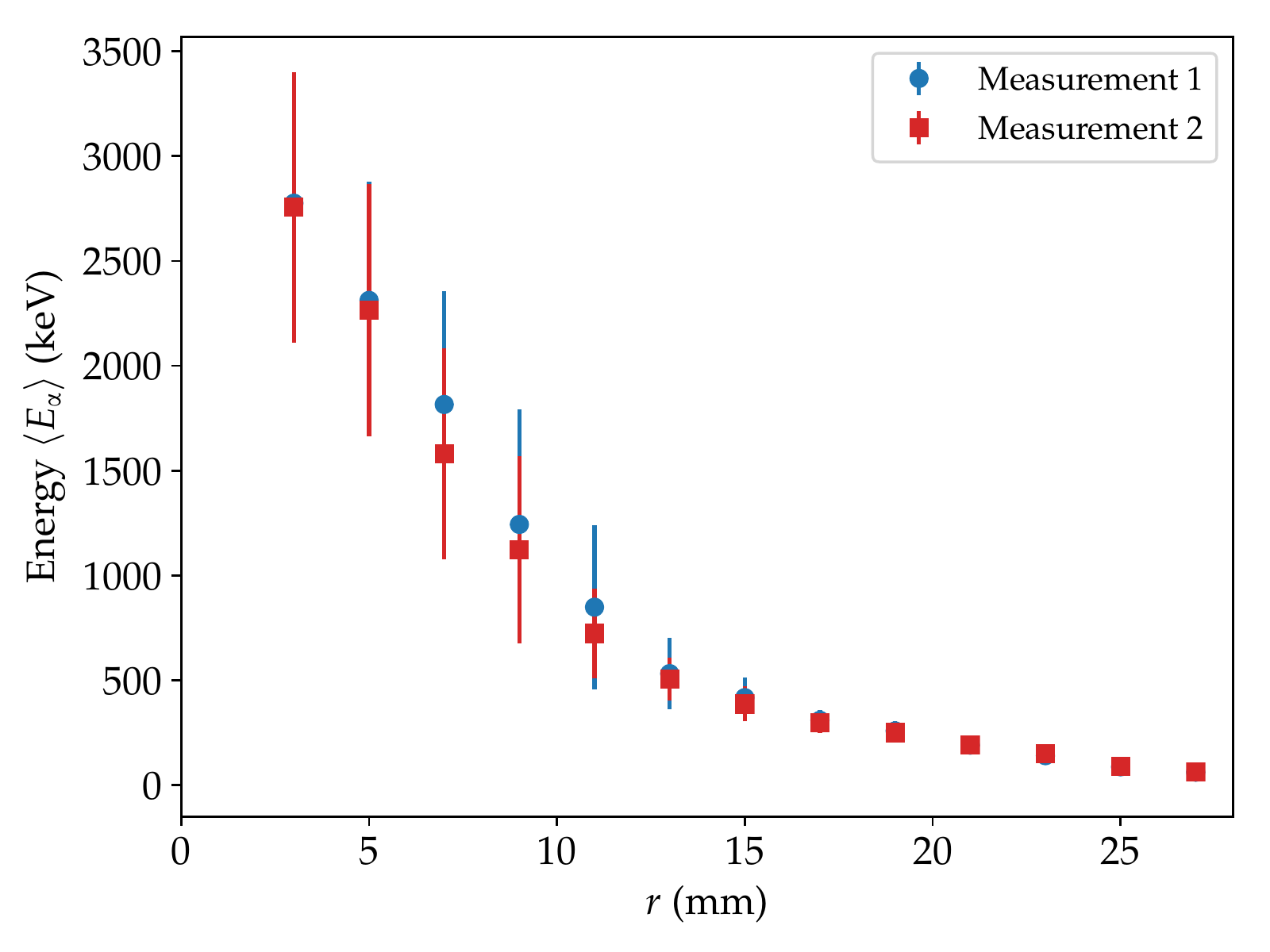}}
\caption{Radial dependence of the observed energy of alpha events: (a)~Energy spectra of the alpha populations at selected radial positions after the application of multivariate cuts. (b)~Mean observed alpha energy $\braket{E^{\text{obs}}_{\alpha}}$ as a function of radius~$r$ for two radial $^{241}$Am scans at different azimuthal positions. Error bars represent the standard deviations of the constrained $E^{\text{obs}}_{\alpha}$ distributions. The measurements were conducted with the detector operated at a bias voltage of $V_{\text{B}}=1050\,$V.}
\label{graph:galatea_am241_energy_radial_dep}
\end{figure}

\paragraph{\emph{Dependence of the observed energy on the radius}}
To quantify the dependence of the observed energy~$E^{\text{obs}}_{\alpha}$ on the radius~$r$ for surface alpha events, the energies of the events in the alpha-enriched regions were histogrammed and corrected for background events, see \textbf{Fig.}~\ref{graph:galatea_am241_alpha_energy_histo}. At small~$r$, the alpha events form a broad distribution with relatively high $E^{\text{obs}}_{\alpha}$. As $r$ increases, the distribution becomes narrower and shifts to lower values~$E^{\text{obs}}_{\alpha}$. A quantitative description of this degradation was obtained by extracting the mean alpha energies~$\braket{E^{\text{obs}}_{\alpha}}$ from the distributions. The ranges of $E^{\text{obs}}_{\alpha}$ were constrained manually to reject remaining background events. The mean alpha energies~$\braket{E^{\text{obs}}_{\alpha}}$ as a function of~$r$ for two radial $^{241}$Am scans at different azimuthal positions are shown in \textbf{Fig.}~\ref{graph:galatea_am241_alpha_radial_energy}. At the outermost radii, $\braket{E^{\text{obs}}_{\alpha}}$ is strongly reduced, i.e.~almost no charges are collected. In contrast, at small~$r$, the mean energy is~$\braket{E^{\text{obs}}_{\alpha}}>2500\,$keV. The results of the two measurements are in good agreement. The observed reduction of the mean alpha energy is consistent with the presence of negative surface charges on the passivation layer, cf.~\textbf{Ch.}~\ref{ch:surface_effects}. These charges trap the holes created during the interaction, reducing the signal amplitude. The broadness of the peak at low~$r$ is partly influenced by the size of the beam spot. However, the total width also includes some stochastic effect. Between the two scans, the detector was unbiased and the cryostat was re-evacuated. This did not affect the observations significantly.
%\FloatBarrier
\noindent
\begin{figure}[!h]
\begin{center}
%\hspace{-2cm}
\includegraphics[angle=0,width=0.6\linewidth]{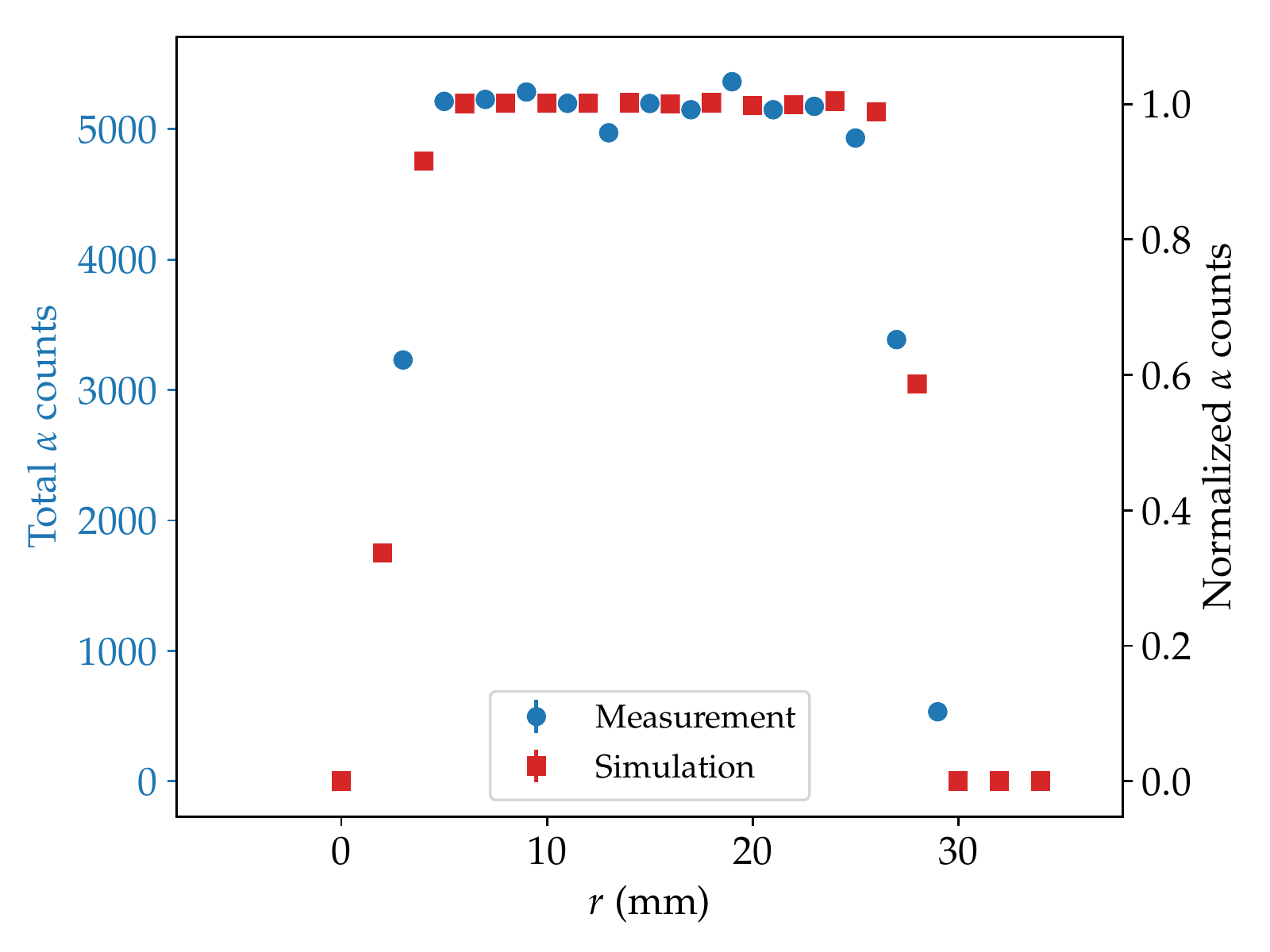}
\caption{Dependence of the number of recorded alpha events on the radius~$r$. Predictions from \textsc{Geant4} simulations are also shown.}
\label{graph:galatea_am241_alpha_rates}
\end{center}
\end{figure}
\FloatBarrier
\noindent
Even though $\braket{E^{\text{obs}}_{\alpha}}$ strongly depends on~$r$, alpha events were not lost. This could have occurred for a locally increasing dead layer. However, the total number of alpha events as a function of $r$ remained almost constant at a value of about $5000$~events per scan position, see \textbf{Fig.}~\ref{graph:galatea_am241_alpha_rates}. The deviation at small radii is most likely due to a partial shadowing of the ${}^{241}$Am beam spot by the PTFE~bar. The decreasing alpha rate at the outer radii can be explained by the fact that the alphas hit the lithiated layer of the taper which they cannot penetrate. The plot also shows the alpha counts as predicted by \textsc{Geant4} simulations. They agree very well with the measurement. The plateau as well as the drop of the event rate at the center and at the edge of the detector are well described by the simulations. However, a small offset of~$3\,$mm was observed. This was found to be due to a corresponding slight offset between the center of the detector and the central position of the collimator. The $r$~values for the data were corrected for this offset.

\begin{figure}[!h]
%\myfloatalign
\subfloat[DCR rate.] {\label{graph:galatea_am241_dcr_rate} 
\includegraphics[width=.48\linewidth]{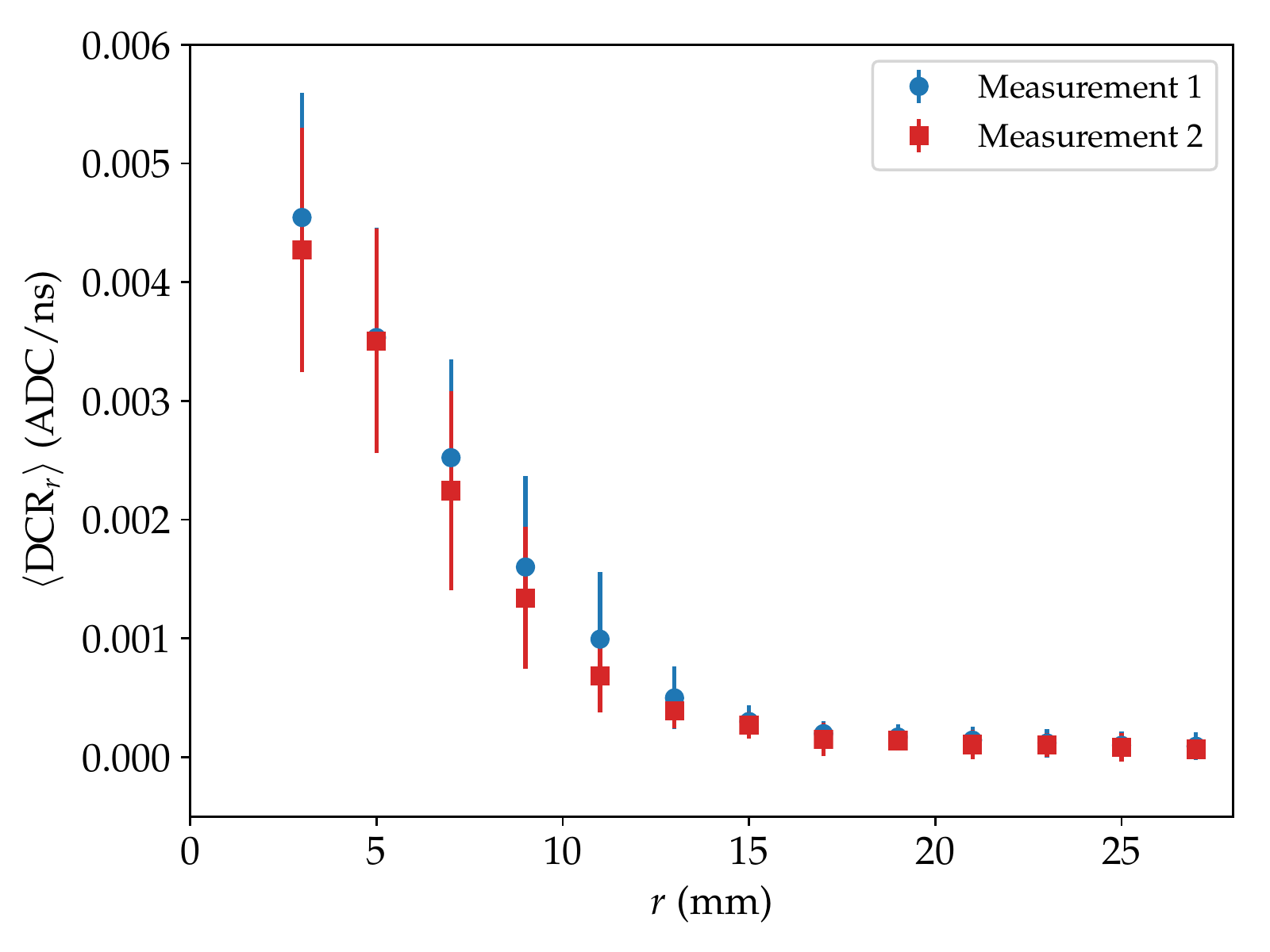}}\quad
\subfloat[DCR fraction.] {\label{graph:galatea_am241_dcr_fraction} 
\includegraphics[width=.48\linewidth]{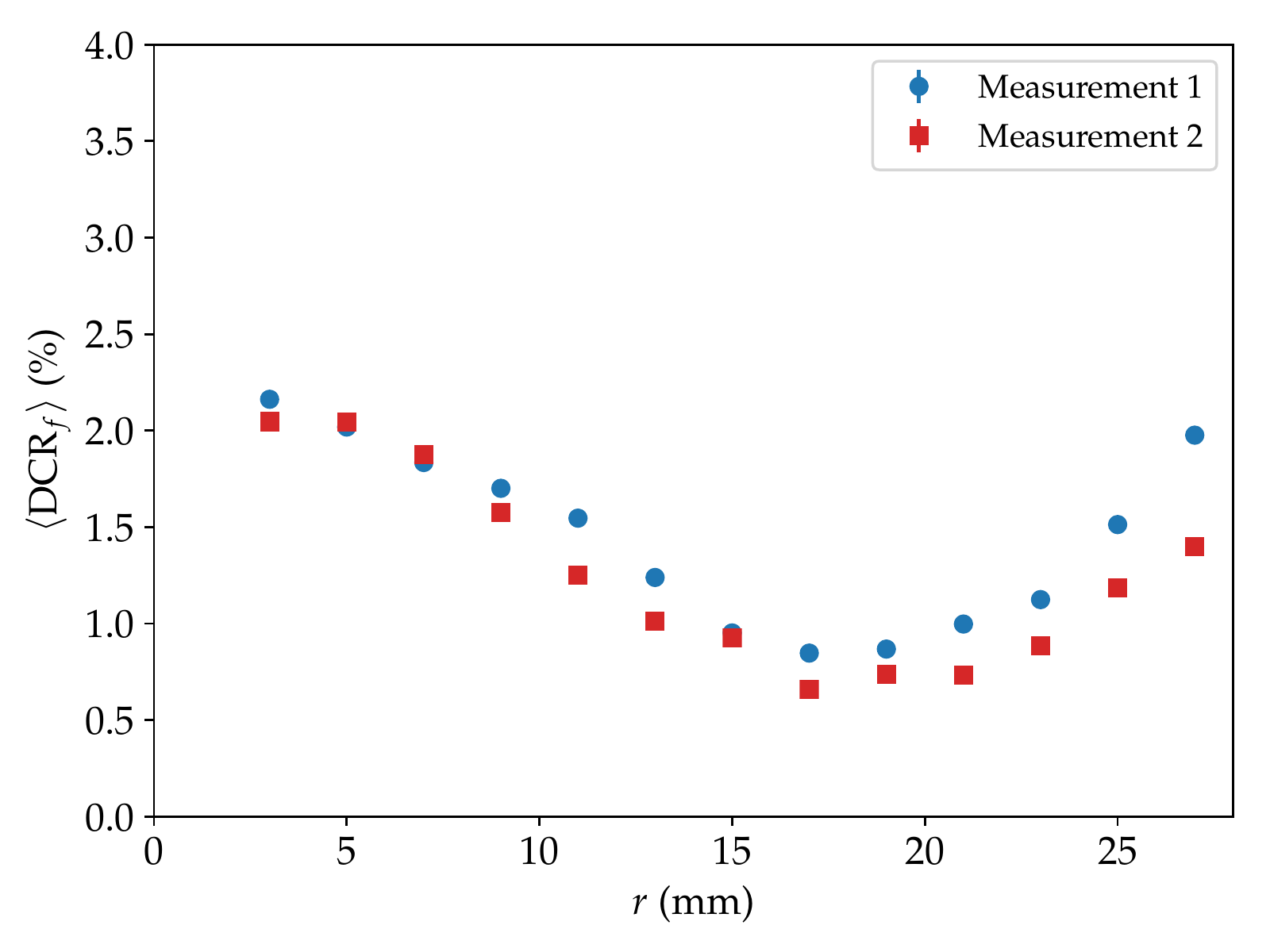}}
\caption{Radial dependence of (a)~$\braket{\text{DCR}_{\text{r}}}$ and (b)~$\braket{\text{DCR}_{\text{f}}}$ for two radial $^{241}$Am scans at different azimuthal positions.}
\label{graph:galatea_am241_dcr_rate_frac}
\end{figure}

\paragraph{\emph{Radial DCR dependence}}
The DCR effect is, in general, an effective way to identify surface alpha events. To investigate its radial dependence, the DCR rate parameter was computed for every event as described in \textbf{Ch.}~\ref{ch:dcr}. The DCR rates were then histogrammed, corrected for background events, and the mean DCR rates~$\braket{\text{DCR}_{\text{r}}}$ were extracted from the distributions. The dependence of $\braket{\text{DCR}_{\text{r}}}$ on $r$ for two radial $^{241}$Am scans at different azimuthal positions is shown in \textbf{Fig.}~\ref{graph:galatea_am241_dcr_rate}. Comparable to the mean alpha energy~$\braket{E^{\text{obs}}_{\alpha}}$, $\braket{\text{DCR}_{\text{r}}}$ decreases considerably with increasing~$r$. At~$r>15$\,mm, the mean DCR rate of surface alpha events is close to zero. This means that in this region alpha events are no longer distinguishable from bulk events.
\FloatBarrier
\noindent
Another way of quantifying the DCR~effect is to convert the mean DCR rate~$\braket{\text{DCR}_{\text{r}}}$ to an average DCR~fraction~$\braket{\text{DCR}_{\text{f}}}$, which is defined as $\braket{E^{\text{ex}}_{\alpha}} / \braket{E^{\text{obs}}_{\alpha}}$, where $\braket{E^{\text{ex}}_{\alpha}}$ is the amount of additionally observed energy due to DCR. This extra observed energy is calculated by converting the mean DCR~rate from ADC/ns to keV/ns units and integrating over the length of the waveform tail $\Delta t \approx14\,\text{\textmu s}$. The radial dependence of~$\braket{\text{DCR}_{\text{f}}}$ is shown in~\textbf{Fig.}~\ref{graph:galatea_am241_dcr_fraction}. The value of $\braket{\text{DCR}_{\text{f}}}$ drops with increasing $r$ from about 2\,\% to about 0.5\,\% at $r \approx 15$\,mm, where the value of the mean DCR rate approaches zero. At higher $r$, $\braket{\text{DCR}_{\text{f}}}$ seems to increase again. However, it should be noted that these fractions are numerically problematic as also $\braket{E^{\text{obs}}_{\alpha}}$ approaches zero, see \textbf{Fig.}~\ref{graph:galatea_am241_alpha_radial_energy}.

%#############################################################################
\subsection{Characterization of surface beta interactions}

\subsubsection{Source configuration}
For the surface characterization measurements with beta particles, a $^{90}$Sr source with an activity of~$A_0=5.0\,$MBq was mounted in the top collimator. Suitable cylindrical tungsten segments were used to fill the collimator frame. Based on the collimator geometry and the source strength, an electron rate of ${\sim}\,300~\text{counts/s}$ was expected at the detector surface. In all measurements, the $^{90}$Sr beam spot had an incidence of~$90^{\circ}$ on the detector surface.
\\\\
As for the surface characterization measurements with alpha particles, several radial scans at different azimuthal positions, as well as background and stability measurements were conducted. In contrast to the alpha measurement configuration, no radial offset between the center of the detector and the central position of the collimator was observed for the beta measurements: The detector holding structure and the collimator were readjusted between the measurement campaigns. Typically, a measurement time of~$0.5\,$hr per scan point was chosen. The measurements were conducted with the detector operated either at the bias voltage of $V_{\text{B}}=1050\,$V or at $V_{\text{B}}=2000\,$V. The focus of the analysis is on the data obtained at the higher bias voltage. Less pronounced results were obtained for the data taken with the lower bias voltage. The observed small dependence on the bias voltage is not yet fully understood. 

\subsubsection{Results}\label{ch:sr90_radial_dependence}

\begin{figure}[!h]
\begin{center}
\includegraphics[width=.6\linewidth]{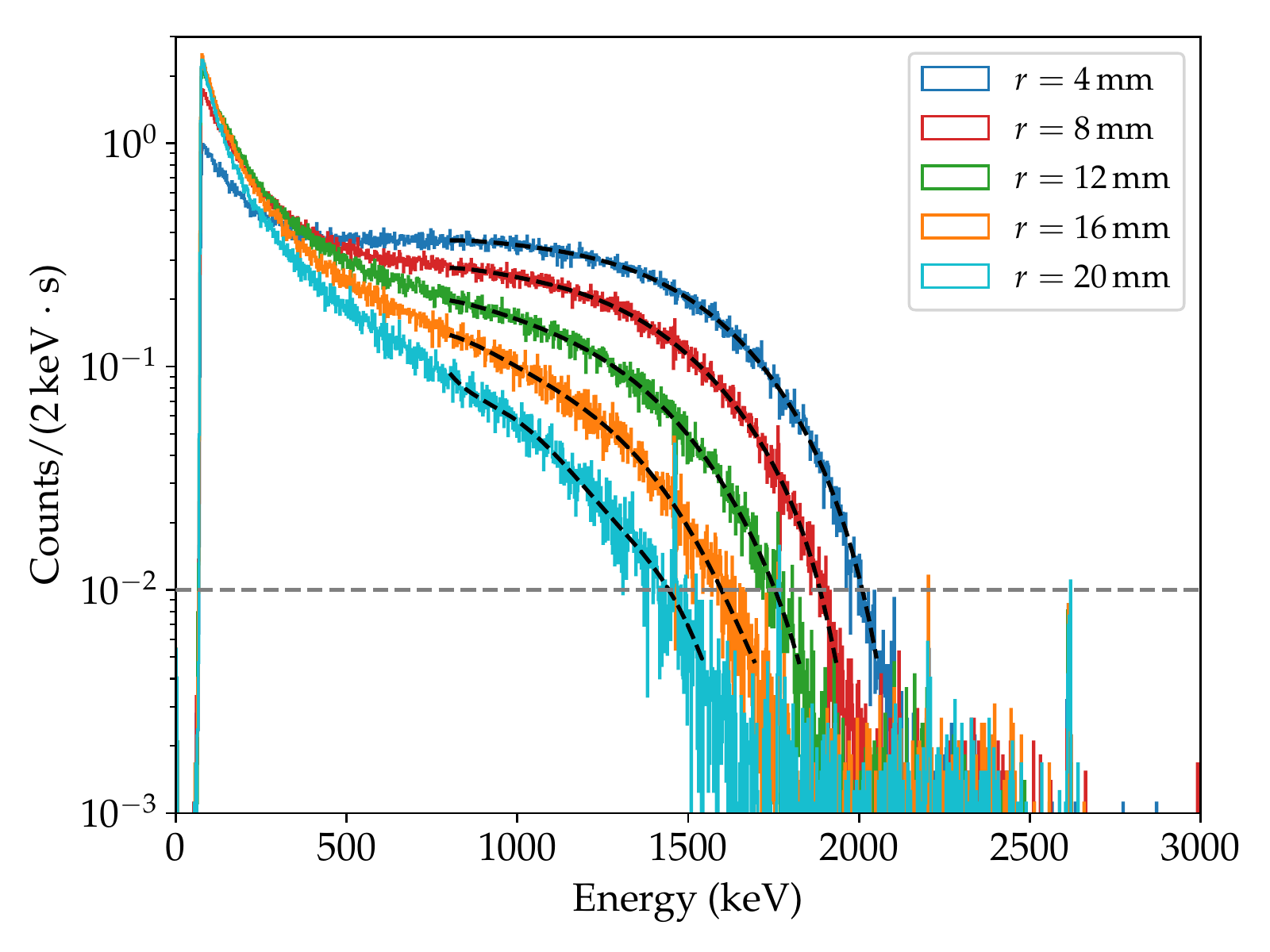}
\caption{Background-subtracted energy spectra~$E^{\text{obs}}_{\beta}$ of a radial $^{90}$Sr scan. The black dashed lines correspond to polynomial fits to the spectra. The endpoint was approximated by determining the intersection of the fits with a fixed value~(grey dashed line).}
\label{graph:galatea_sr90_bkg_corr_spectra}
\end{center}
\end{figure}

\begin{figure}[!h]
%\myfloatalign
\subfloat[Integral count rate.] {\label{graph:galatea_sr90_rate_evolution} 
\includegraphics[width=.48\linewidth]{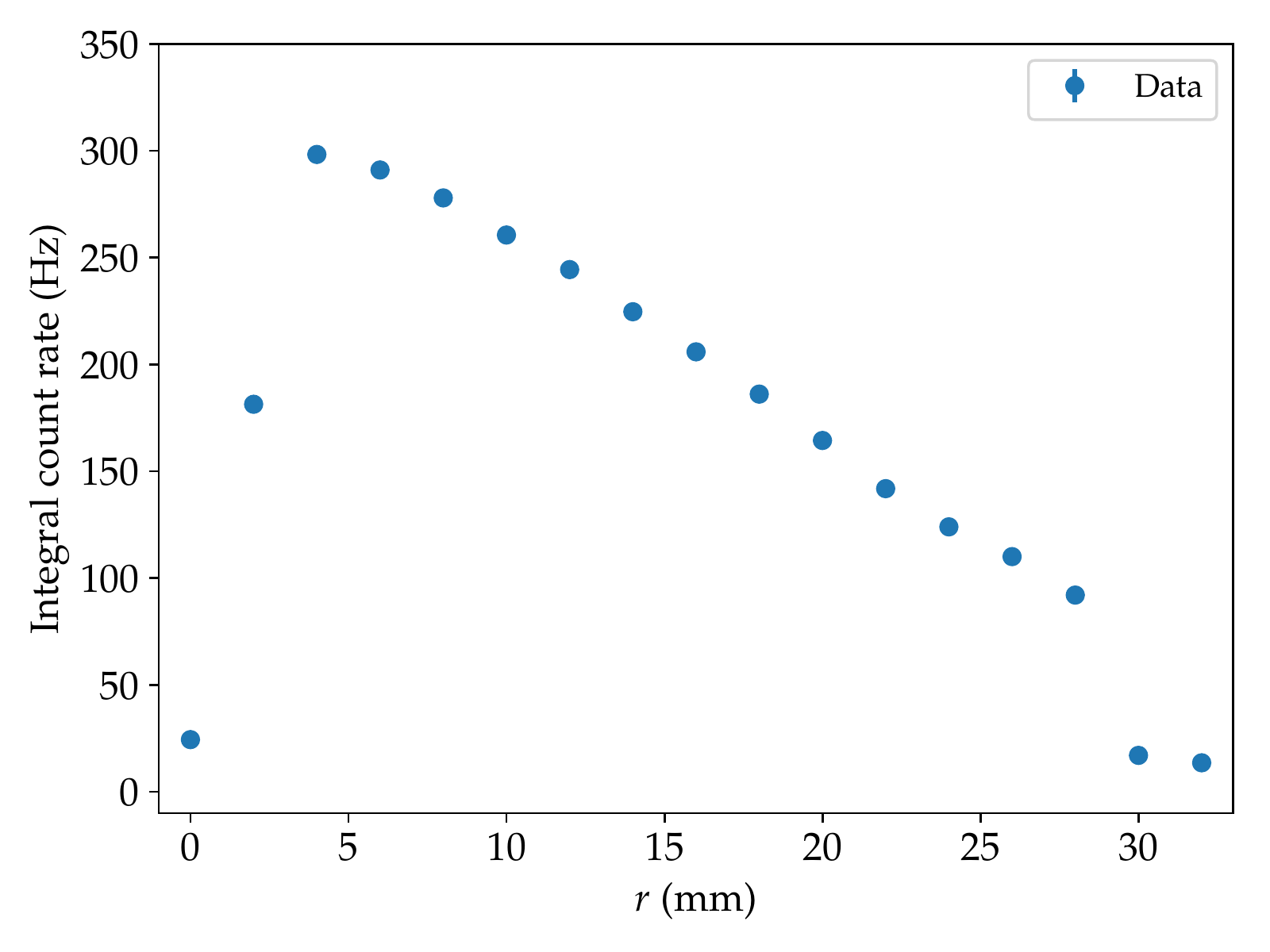}} \quad
\subfloat[Relative endpoints.] {\label{graph:galatea_sr90_rel_endpoints}
\includegraphics[width=.48\linewidth]{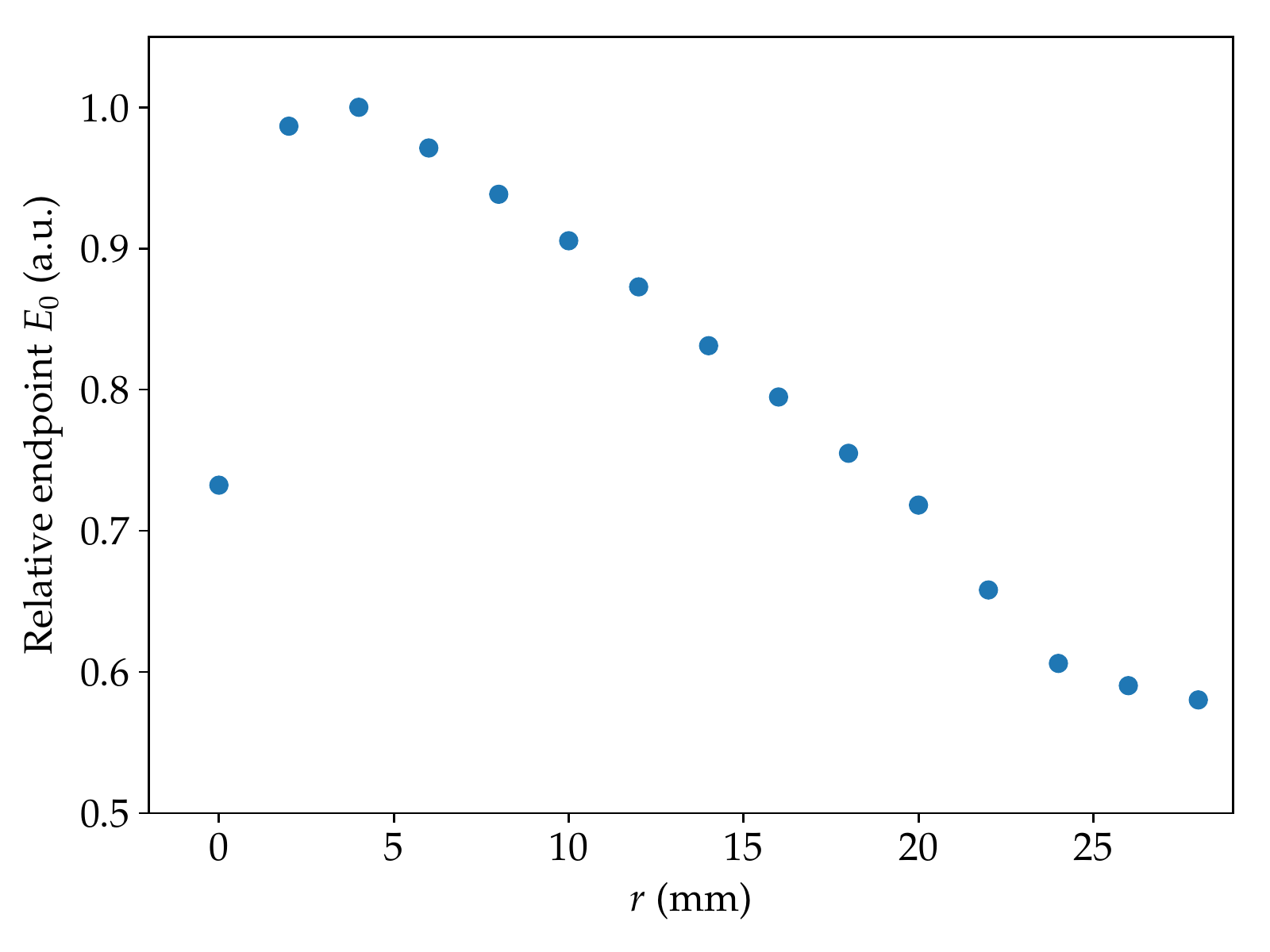}}
\caption{(a)~Dependence of the integral count rate and (b)~of the relative endpoint on $r$ as observed for a $^{90}$Sr scan. The measurements were conducted with the detector operated at a bias voltage of $V_{\text{B}}=2000\,$V.} 
\label{graph:galatea_sr90_rate_endpoint}
\end{figure}

\paragraph{\emph{Dependence of the observed energy on the radius}}
First, the energy spectra recorded in the presence of the $^{90}$Sr source were corrected for background events, see \textbf{Fig.}~\ref{graph:galatea_sr90_bkg_corr_spectra}. The plot shows that the distribution of~$E^{\text{obs}}_{\beta}$ strongly depends on the radius~$r$. In particular, the following two effects can be observed:
\begin{enumerate}
\item[$1)$] The total number of events decreases with increasing radius~$r$.
\item[$2)$] The energy continuum $E^{\text{obs}}_{\beta}$ shifts to lower energies with increasing $r$. This is especially pronounced around the endpoint of the distribution.
\end{enumerate}
\FloatBarrier
\noindent
The first effect was quantified by calculating the total count rate as integrated over the entire energy range~($0-3\,$MeV) as a function of $r$, see \textbf{Fig.}~\ref{graph:galatea_sr90_rate_evolution}. The count rate first increases and then decreases with~$r$. The reduced rate at small $r$ is due to the partial shadowing of the beam spot by the PTFE bar. While the event rate for alpha events was almost constant, the reduced event rate at higher radii shows that some of the beta electrons are completely lost. This is most likely an experimental artefact: As the activity of the $^{90}$Sr source was higher than the activity of the $^{241}$Am source, the trigger threshold of the data acquisition system had to be increased from~${\sim}\,16\,$keV to ${\sim}\,50\,$keV to prevent too high a rate of pile-up events. Therefore, events which were affected severely by surface effects and thus with too small $E^{\text{obs}}_{\beta}$ were not recorded.
\\\\
The dependence of the spectral endpoint on~$r$ was investigated by fitting the energy spectra with a seventh-order polynomial, see \textbf{Fig.}~\ref{graph:galatea_sr90_bkg_corr_spectra}. The endpoints~$E^{\text{obs}}_0$ were approximated by determining the energies, at which the fit functions drop to a fixed value of~$10^{-2}~\text{counts}/(2\,\text{keV}\cdot\text{s})$. This value was chosen to avoid statistical fluctuations at smaller count rates. As the estimate of the endpoints is rough and since this method is affected by binning effects, normalized endpoints $E_0=E^{\text{obs}}_0/\text{max}(E^{\text{obs}}_0)$ are shown in \textbf{Fig.}~\ref{graph:galatea_sr90_rel_endpoints}. The value of $E_0$ decreases significantly with~$r$. This is in good qualitative agreement with the behavior of $E^{\text{obs}}_{\alpha}$ as described in~\textbf{Ch.}~\ref{ch:alpha_characterization}. In particular, the results are again consistent with the presence of negative charges on the passivated detector surface.
\FloatBarrier
\noindent

\begin{figure}[!h]
\begin{center}
%\hspace{-2cm}
\includegraphics[angle=0,width=0.73\textwidth]{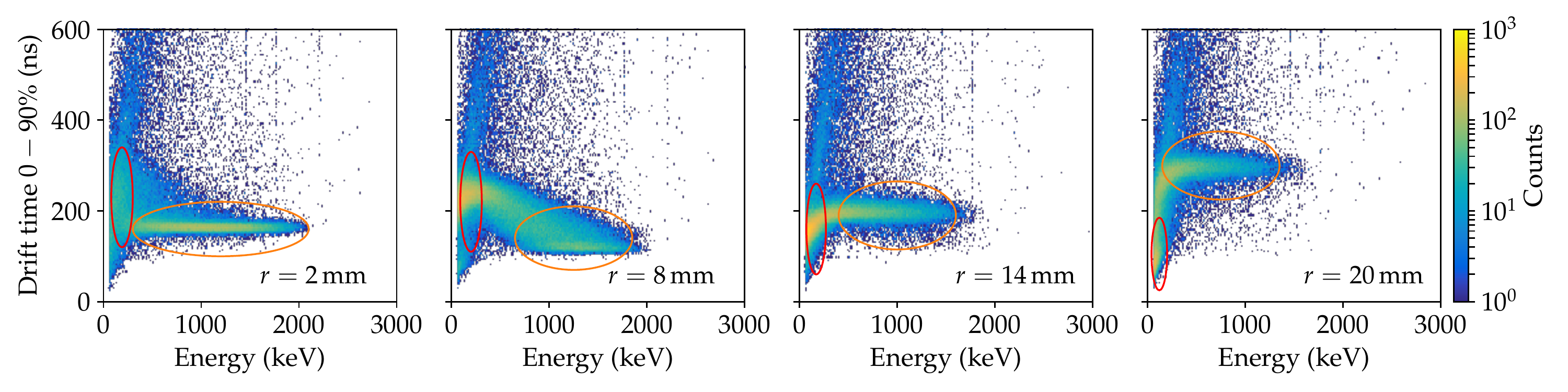}
\caption{Correlations between the drift time ($0-90\%$) and $E^{\text{obs}}_{\beta}$ at selected $r$. The red vertical and orange horizontal ellipses in the distributions indicate two event populations.}
\label{graph:galatea_sr90_before_cuts}
\end{center}
\end{figure}

\paragraph{\emph{Radial dependence of other pulse shape parameters}}
The radial dependence of other pulse shape parameters and their correlations were also investigated.  Two event populations were identified by studying the correlation between the drift time and the energy~$E^{\text{obs}}_{\beta}$. The drift time is defined as the time period in which~$90\%$ of the total signal height is reached. The correlations for selected radii are shown in \textbf{Fig.}~\ref{graph:galatea_sr90_before_cuts}. 
\FloatBarrier
\noindent
One population, indicated by vertical ellipses, is located at small energies and its drift times decrease with increasing~$r$. The second population, indicated by horizontal ellipses, is located at higher energies and its drift times increase with~$r$. With the help of pulse shape simulations it can be shown that the first population corresponds to events with a small penetration depth, which are sensitive to surface effects. For the observed negative charges on the passivated detector surface, the signal development is driven by the collection of electrons while the holes are almost stationary and provide an almost constant contribution. Since at higher $r$ the electrons are closer to the n$^+$~contact, their drift time decreases. As the weighting potential at higher $r$ is small, $E^{\text{obs}}_{\beta}$ is small for these events. The second population corresponds to events with higher penetration depths which are mostly insensitive to surface effects. These interactions are subject to the usual charge collection behavior, i.e.~the holes are at least partially collected and $E^{\text{obs}}_{\beta}$ is closer to the true energy of the incident electrons. Since at higher $r$, the holes have a longer drift path to the p$^+$ readout contact, the drift time increases with~$r$. These separated two event populations show that, unlike alphas, not all electrons are affected significantly by surface charges. Only events where the electrons do not penetrate deeply are strongly affected.

%#############################################################################
%#############################################################################
%#############################################################################
\section{Pulse shape simulations}\label{ch:simulations}
To better understand the measurement results discussed above, dedicated surface event simulations were performed. To this end, the package \texttt{Siggen} consisting of the two programs \texttt{mjd\_fieldgen} and \texttt{mjd\_siggen} was used~\cite{radford_siggen}. 

\paragraph{\emph{Electric field and weighting potential}}
The stand-alone program \texttt{mjd\_fieldgen} was used to calculate the electric potential, the electric field, and the weighting potential inside the detector. The computation is based on a numerical relaxation algorithm on an adaptive grid. For PPC~detectors, due to their cylindrical symmetry, the computation can be performed on a two-dimensional grid (coordinates: $r$ and $z$). At the passivated detector surface, a reflective symmetry is used as a boundary condition for the relaxation algorithm. This is in accordance with the requirement that for zero surface charge at the passivation layer, the field lines close to the surface are parallel to that surface, such that no charges pass the surface~\cite{mertens2019}.

\paragraph{\emph{Signal formation}}
The signals corresponding to energy depositions at specific locations in the detector can be simulated with \texttt{mjd\_siggen}. The program combines the field maps generated with \texttt{mjd\_fieldgen} with a charge drift model containing information on the electron and hole mobilities to compute the charge drift path~\cite{mullowney2012}. Furthermore, the corresponding signal is calculated according to the Shockley-Ramo theorem, cf.~\textbf{Eq.}~(\ref{eq:shockley_ramo}).
\\\\
Both programs require a number of user inputs that are read in from a common configuration file. These inputs include the detector geometry and configuration (bias voltage, temperature), the impurity profile, and other settings (initial grid size, charge cloud size, electronics response, etc.). Most importantly for this work, a (homogeneously distributed) surface charge can be added to the passivated detector surface. The surface charge is expressed in units of~e/cm$^2$ and is added as an impurity at every grid point on the surface.

%%%%%%%%%%%%%%%%%%%%%%%%%%%%%%%%%%%%%%%%%%%%%%%%%%%%%%%%%%%%%%%%%%%%%%%%%%%%%%%%%%%%%%%%%%%%%
\subsection{Influence of surface effects on pulse shape parameters}\label{ch:charge_efficiency_maps}
\begin{figure}[!h]
%\myfloatalign
\subfloat[Energy fraction, $\sigma<0$.]{\includegraphics[width=.33\linewidth]{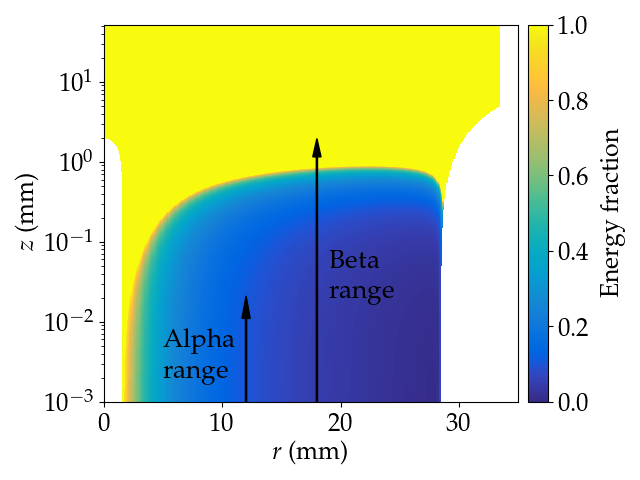}}%\quad
\subfloat[$A/E$, $\sigma<0$.]{\includegraphics[width=.33\linewidth]{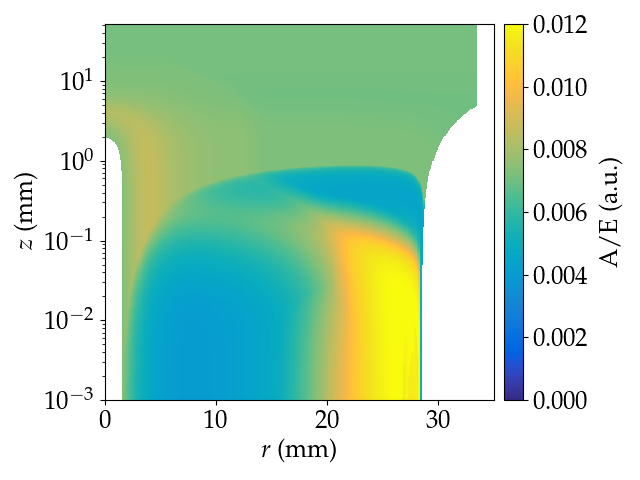}}%
\subfloat[Drift time ($0-90\%$), $\sigma<0$.]{\includegraphics[width=.33\linewidth]{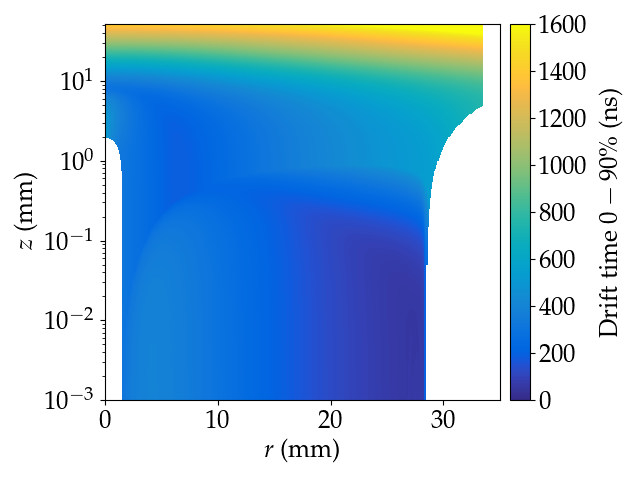}}\\
\subfloat[$E^{\text{obs}}/E^{\text{true}}$, $\sigma>0$.] {\label{graph:galatea_simu_psm_energy_frac}
\includegraphics[width=.33\linewidth]{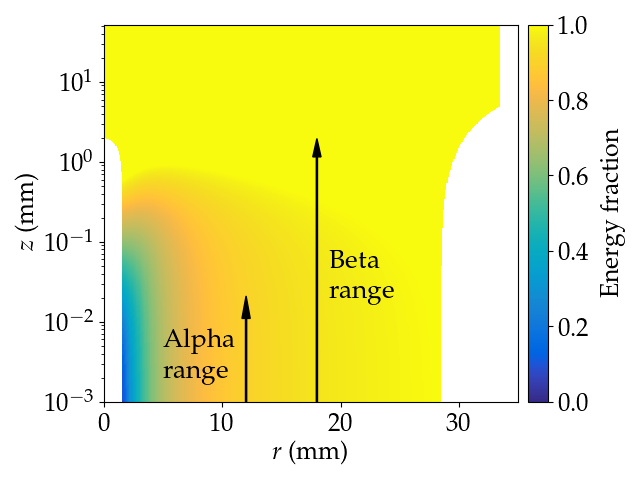}}%\quad
\subfloat[$A/E$, $\sigma>0$.] {\label{graph:galatea_simu_psm_aoe} 
\includegraphics[width=.33\linewidth]{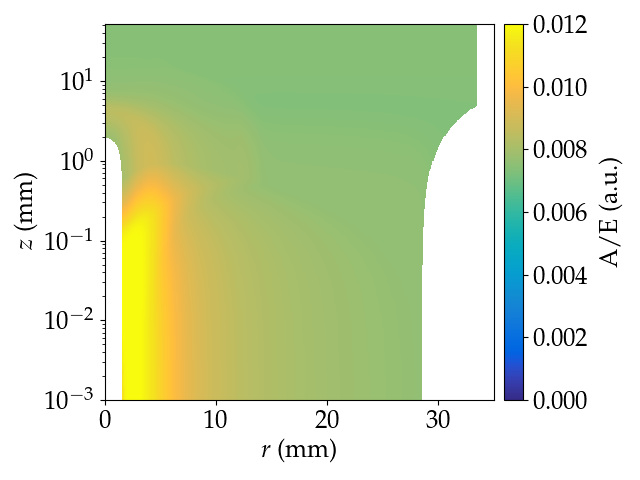}}%\\
\subfloat[Drift time ($0-90\%$), $\sigma>0$.] {\label{graph:galatea_simu_psm_drifttime}
\includegraphics[width=.33\linewidth]{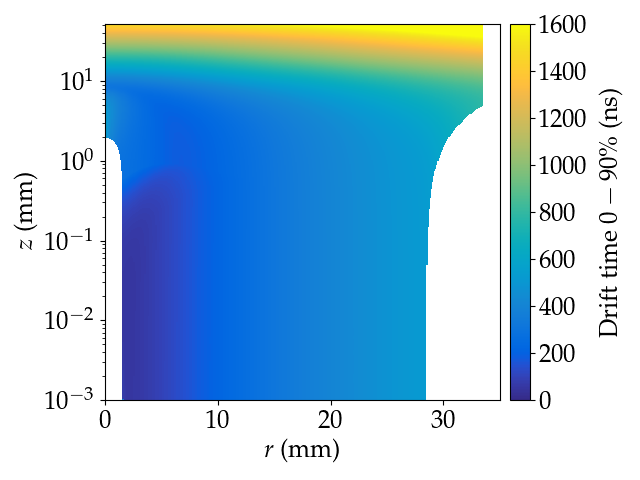}}\\%\quad
\caption{Pulse shape parameter maps of the quantities energy fraction~$E^{\text{obs}}/E^{\text{true}}$, $A/E$, and drift time~($0-90\%$) for a negative surface charge (\mbox{$\sigma=-0.3\cdot10^{10}\,\text{e}/\text{cm}^2$}, top row), and for a positive surface charge (\mbox{$\sigma=+0.3\cdot10^{10}\,\text{e}/\text{cm}^2$}, bottom row) at a bias voltage of~$V_{\text{B}}=1050\,$V. The distributions are shown with a logarithmic scale in~$z$ to highlight surface effects at the passivation layer. The typical ranges of alpha and beta particles within the active detection volume are indicated by the arrows.}
\label{graph:galatea_simu_pulse_shape_maps}
\end{figure}

Pulse shape parameter maps were calculated to study the impact of surface effects on important quantities such as the energy~$E^{\text{obs}}$, drift time, etc. To this end, point charges with starting positions arranged in a finely meshed grid in the $(r,z)$~plane were simulated using \texttt{Siggen}. The effects of diffusion and self-repulsion were not included in the simulations, cf.~\textbf{Ch.}~\ref{ch:simu_diffusion_repulsion}. The parameter maps for negative and positive surface charges ($\sigma=\pm0.3\cdot10^{10}\,\text{e}/\text{cm}^2$) for the quantities energy fraction $E^{\text{obs}}/E^{\text{true}}$, $A/E$, and drift time~($0-90\%$) are shown in \textbf{Fig.}~\ref{graph:galatea_simu_pulse_shape_maps}. The $A/E$~parameter describes the ratio of the maximum amplitude of the current pulse~($A$), and the amplitude (energy) of the charge pulse~($E^{\text{obs}}$). It is commonly used to discriminate background events from signal events. More information on this pulse shape parameter and its determination can be found in~\cite{agostini2013,wagner2017}.
\\\\
The energy fraction parameter maps for negative and positive surface charges show that in most of the active detection volume, the true event energy is obtained, i.e.~$E^{\text{obs}}\approx E^{\text{true}}$. However, for $\sigma<0$, there is a strong reduction of $E^{\text{obs}}$ in a region close to the passivated surface~($z\lesssim1\,$mm). In this region, holes created during an interaction are attracted to the surface and become quasi-stationary. The signal development is driven by the drift of electrons to the n$^+$~contact, cf.~\textbf{Ch.}~\ref{ch:surface_effects}. At~$z\lesssim1\,$mm, the energy fraction~$E^{\text{obs}}/E^{\text{true}}$ decreases for increasing~$r$. Surface alpha events with typical penetration depths of tens of micrometers are fully contained in this region of reduced~$E^{\text{obs}}$. In contrast, surface beta events which have in average higher penetration depths of up to a few millimeters, are only partially affected. 
\FloatBarrier
\noindent
In the case of positive surface charges, $\sigma>0$, there is only a small region in the vicinity of the point contact where events have a strongly reduced~$E^{\text{obs}}$ and thus small $E^{\text{obs}}/E^{\text{true}}$ values. Here, electrons created during an interaction are attracted to the passivated surface and become quasi-stationary. The signal development is driven by the drift of holes to the p$^+$~contact. At $z\lesssim1\,$mm, the reduction of $E^{\text{obs}}$ gets less severe with increasing~$r$. Compared to the case of a negative surface charge build-up, the reduction of $E^{\text{obs}}$ is much less pronounced. For increasing absolute surface charge $|\sigma|$ (at fixed bias voltage) or decreasing bias voltage (at fixed surface charge), the regions of reduced $E^{\text{obs}}$ extend towards higher depths~$z$.
\\\\
The $A/E$~map for a negative surface charge shows that for~$z\lesssim1\,$mm, the $A/E$~values first slightly decrease and then strongly increase with increasing $r$. This can be explained by the fact that at larger $r$, the electrons drift in a slowly changing weighting field for a short drift time which results in fast signals and thus high $A/E$~values. For positive surface charges, high $A/E$~values are encountered in the region close to the point contact. Here, the holes drift in a rapidly changing weighting field for a short drift time which also results in fast signals.
\\\\
The drift time of events in the region of reduced $E^{\text{obs}}$ for $\sigma<0$ decreases with increasing~$r$. This is due to the closer proximity of the electrons to the n$^+$~electrode at higher~$r$ so they are collected faster. The drift time map also shows that at higher penetration depths~($z\gtrsim1\,$mm), the drift time increases with increasing~$r$. Here, the holes are collected and their drift times drive what is measured. At larger~$r$, they have a longer drift path to the p$^+$~contact and therefore a longer drift time. Likewise, for positive surface charges, where the signal formation in the region of reduced~$E^{\text{obs}}$ is driven by the collection of holes, the drift time increases with radius.

\subsection{Full Monte Carlo simulations}
\begin{figure}[!h]
\begin{center}
\includegraphics[width=.5\linewidth]{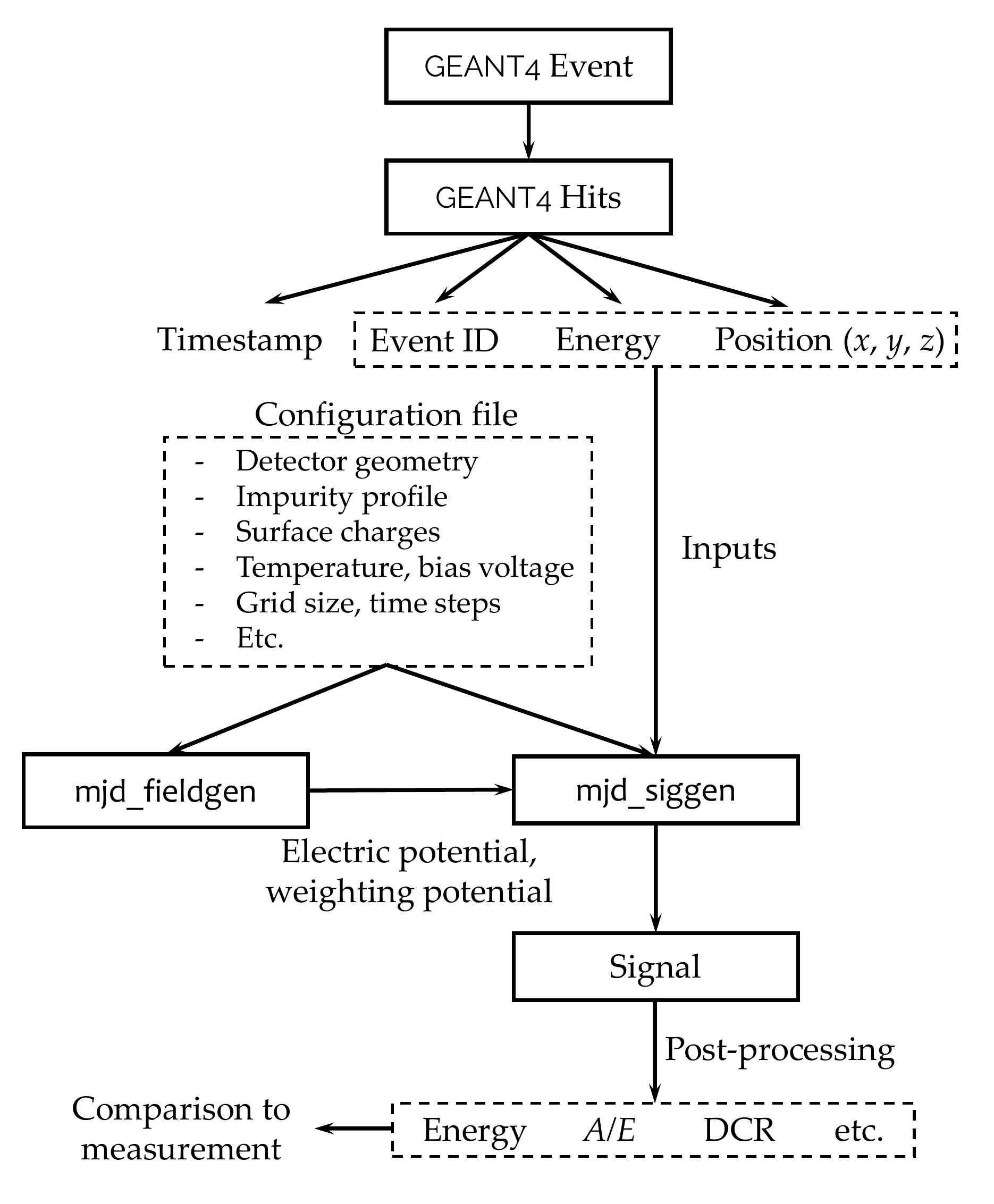}
\caption{Flow diagram of the surface event simulations.}
\label{graph:galatea_simulation_structure}
\end{center}
\end{figure}

An extensive simulation campaign was carried out to better understand the results obtained in the surface characterization measurements with alpha and beta particles. First, realistic energy deposition distributions in the PPC~detector were simulated using the toolkit \textsc{Geant4}. Second, the corresponding signals were simulated using \texttt{Siggen}. Third, various pulse shape parameters were computed and analyzed in post-processing routines. This three-step procedure is illustrated schematically in \textbf{Fig.}~\ref{graph:galatea_simulation_structure}.
%\FloatBarrier
\noindent

%%%%%%%%%%%%%%%%%%%%%%%%%%%%%%%%%%%%%%%
\paragraph{1) \textsc{Geant4} simulations}
In the first step, the interaction positions and energy depositions of surface alpha and beta events in a PPC germanium detector were simulated using \textsc{Geant4}. To this end, a simplified geometry of the \textsc{Galatea} scanning facility was implemented. To acquire sufficiently high statistics, several million events were simulated. For every simulated event, the parameters timestamp, event~ID, energy~$E^{\text{true}}$, and position~($x, y, z$) were stored for every charge deposition~(hit) in the detector.

%%%%%%%%%%%%%%%%%%%%%%%%%%%%%%%%%%%%%%%
\paragraph{2) Pulse shape simulations}
The outputs of the \textsc{Geant4} simulations were used as an input for pulse shape simulations with \texttt{Siggen}. For a given event, the signals corresponding to the individual hits were simulated and finally summed up to form the signal (weighted with the individual~$E^{\text{true}}$). The evolution of the charge cloud size due to diffusion and self-repulsion was neglected in all simulations. This will be discussed in more detail in~\textbf{Ch.}~\ref{ch:simu_diffusion_repulsion}. Moreover, the implementation of a sophisticated electronic response model (e.g.~modeling the electronic noise) was omitted. The simulated waveform of every event was stored for a time period of~$1500\,$ns (starting at $t=t_0=0\,$ns) for time steps of $\Delta t=1\,$ns. This trace length was chosen to minimize the computing time while ensuring the simulation of the full signal for events close to the passivated detector surface.
  
%%%%%%%%%%%%%%%%%%%%%%%%%%%%%%%%%%%%%%%
\paragraph{3) Post-processing}
In post-processing, several pulse shape parameters were extracted from the simulated waveforms. These include the observed energy~$E^{\text{obs}}$, the maximum current amplitude to estimate~$A/E$, the signal drift and rise time, and the DCR~rate. The DCR~effect for surface alpha events was modeled by convolving the current signal with an exponential, followed by a re-integration to obtain the convolved charge signal~$\hat{S}(t)$:
\begin{align}
\hat{S}(t)=C\sum_{t, t^{\prime}=t+1} \left[S(t)-S(t-1)\right]\left(1-\exp\left(\frac{t-t^{\prime}}{\tau}\right)\right).\label{eq:simu_dcr_convolution}
\end{align}
Here, $C$~denotes a factor containing the fraction of charges released into the detector bulk, $S(t)$~the original (non-convolved) signal, and $\tau$~an exponential time constant describing the time scale of charge release. The equation accounts for the fact that the delayed charges are released starting from when the alpha particle penetrates the surface. The DCR~rate defined in this equation is proportional to~$E^{\text{obs}}$. It should be noted here that the DCR model in \textbf{Eq.}~(\ref{eq:simu_dcr_convolution}) was tuned to match the measured effect as presented in this work. Alpha or beta surface events with other topologies, e.g.~different incidence and/or energy, may not be described well.

%#############################################################################
\subsubsection{Surface alpha events}
\begin{figure}[!h]
\begin{center}
%\hspace{-2cm}
\includegraphics[angle=0,width=0.735\textwidth]{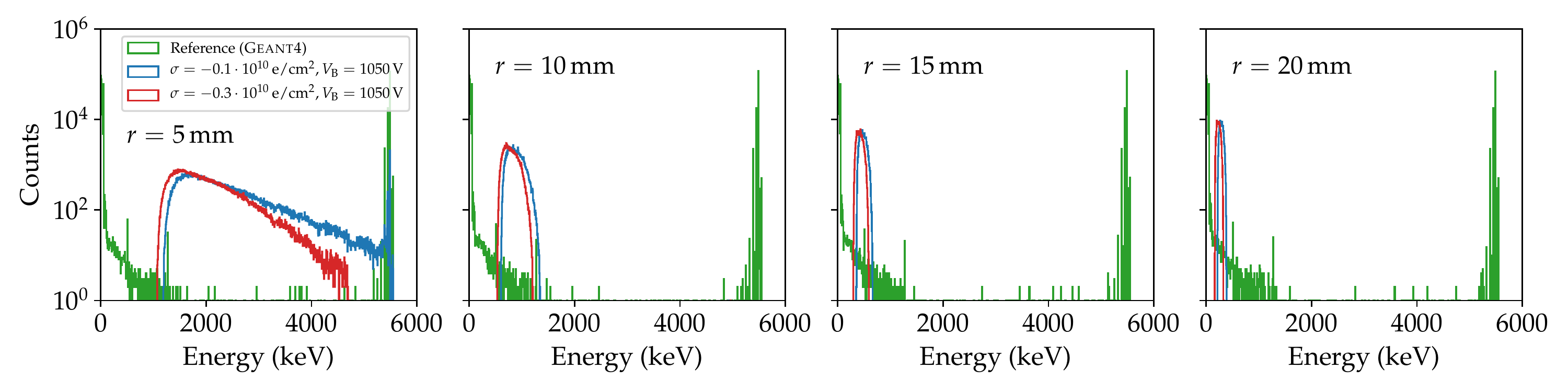}\\
\includegraphics[angle=0,width=0.73\textwidth]{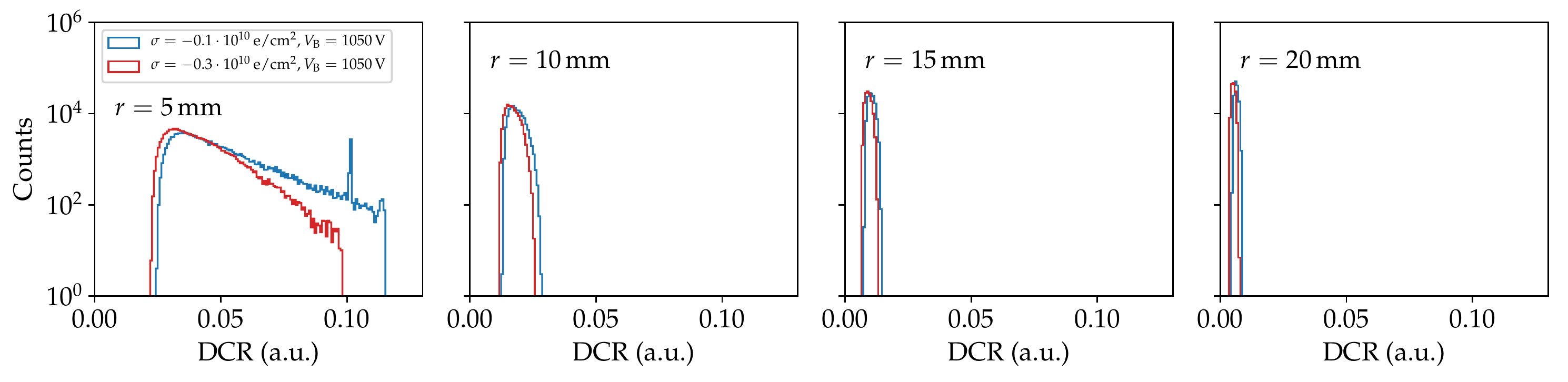}
\caption{Simulated energy spectra $E^{\text{true}}_{\alpha}$ (\textsc{Geant4} reference, green curve), $E^{\text{obs}}_{\alpha}$ (blue and red curves) (upper row), and DCR distributions (lower row) of the $^{241}$Am events at selected~$r$. The simulated~$E^{\text{obs}}_{\alpha}$ distributions are shown for negative surface charges \mbox{$\sigma=-0.1, -0.3\cdot10^{10}\,\text{e}/\text{cm}^2$}.}
\label{graph:galatea_simu_energy_histos_neg}
\end{center}
\end{figure}

\begin{figure}[!h]
%\myfloatalign
\subfloat[Mean observed alpha energies.] {\label{graph:galatea_meas_simu_energy} \hspace{-0.2cm}
\includegraphics[width=.5\linewidth]{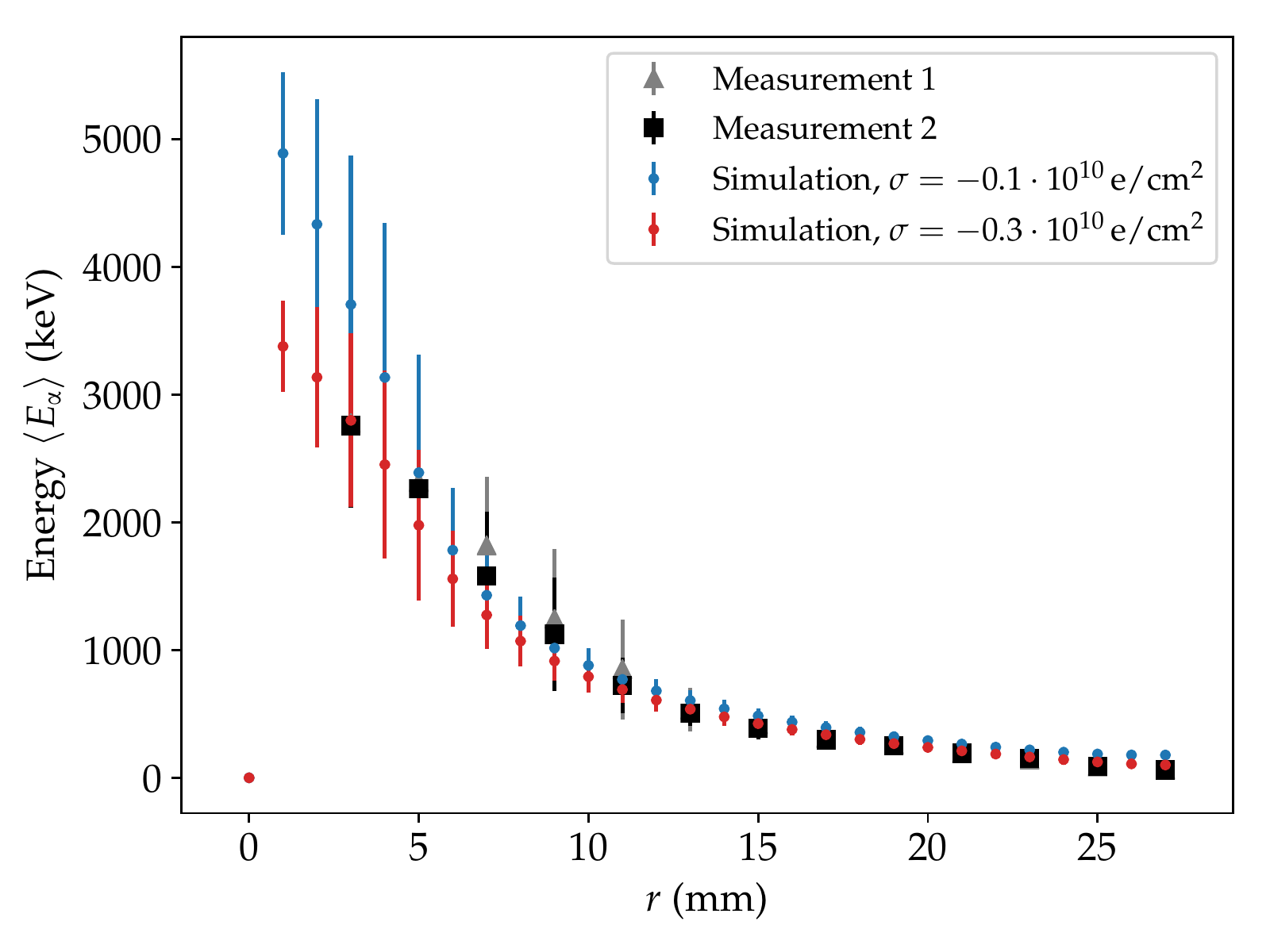}}%\quad
\subfloat[Mean observed DCR~rates.] {\label{graph:galatea_meas_simu_dcr}
\includegraphics[width=.5\linewidth]{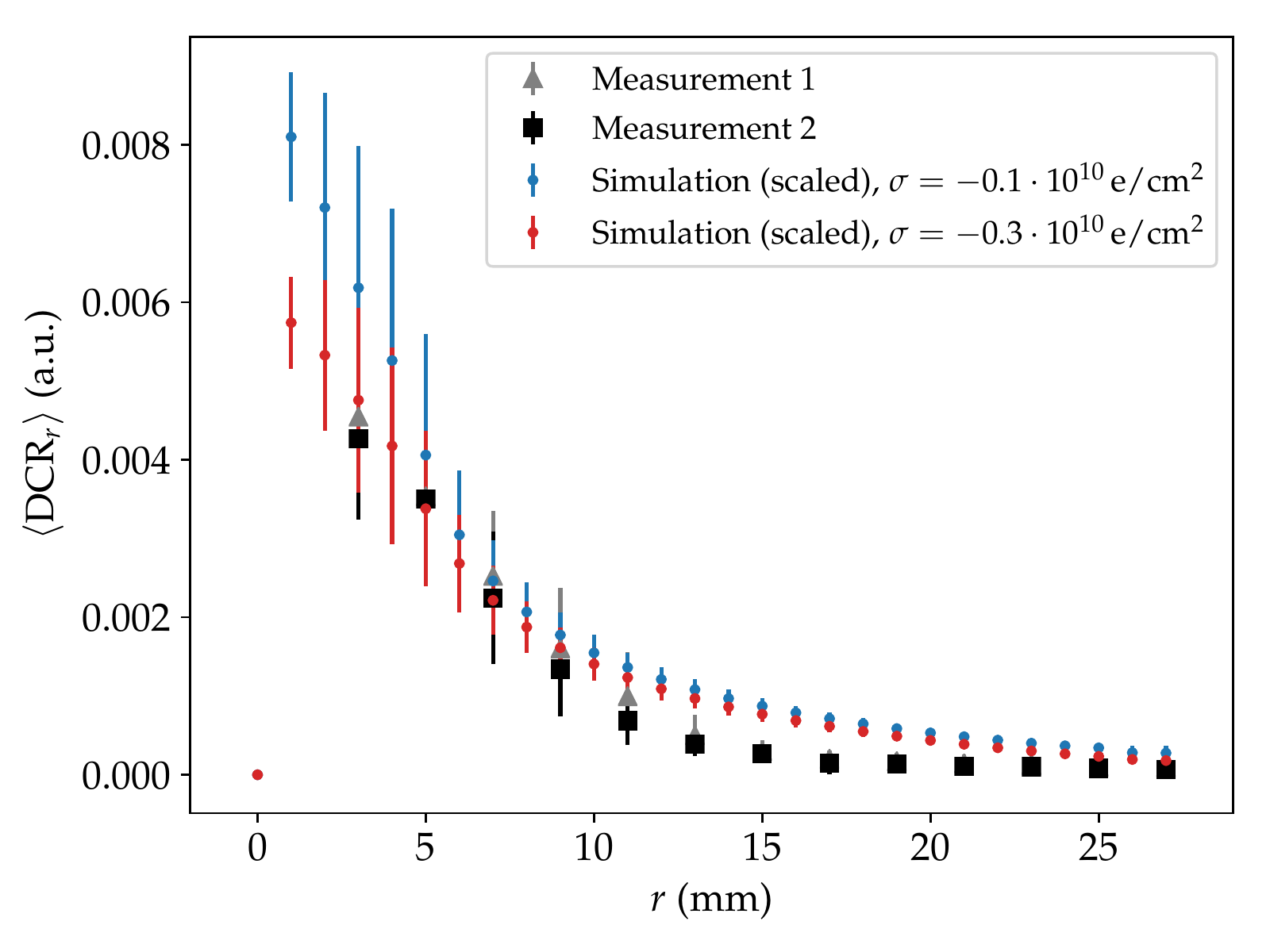}}
\caption{Comparison of measurement to simulation: (a)~mean alpha energy~$\braket{E^{\text{obs}}_{\alpha}}$ and (b)~mean DCR rate $\braket{\text{DCR}_{\text{r}}}$. The simulation results are shown for the surface charges \mbox{$\sigma=-0.1, -0.3\cdot10^{10}\,\text{e}/\text{cm}^2$}. The simulated DCR~rates were scaled with a constant factor.}
\label{graph:galatea_meas_simu_comparison}
\end{figure}

In this section, the results of the $^{241}$Am surface alpha event simulations will be discussed and compared to the  measurements. The energies~$E^{\text{true}}_{\alpha}$, $E^{\text{obs}}_{\alpha}$, and DCR~rates for all simulated events are shown for negative surface charges ($\sigma=-0.1, -0.3\cdot10^{10}\,\text{e}/\text{cm}^2$) in \textbf{Fig}~\ref{graph:galatea_simu_energy_histos_neg}. The simulation predicts that with increasing radius~$r$, the alpha population moves towards lower~$E^{\text{obs}}_{\alpha}$ and DCR values. In addition, the distributions become narrower. This is in good qualitative agreement with the measurements, cf.~\textbf{Fig.}~\ref{graph:galatea_am241_alpha_energy_histo}. However, it should be noted here that there are differences between the spectral shapes of the measured and simulated $^{241}$Am spectra. This is most likely due to the fact that the simulation framework does not fully cover all relevant effects, e.g.~diffusion and self-repulsion of the charge cloud evolution were neglected. Moreover, the simplified simulation settings (e.g.~simplified geometry of the experimental setup, homogeneous distribution of the surface charges, discrete simulation grid, no sophisticated electronics response, etc.) could also have an impact. 
\\\\
To quantify the radial dependencies predicted by the pulse shape simulations, the mean energies~$\braket{E^{\text{obs}}_{\alpha}}$ and the mean DCR~rates~$\braket{\text{DCR}_{\text{r}}}$ were extracted from the distributions, see \textbf{Fig.}~\ref{graph:galatea_meas_simu_comparison}. To eliminate background events (particularly the $59.5\,$keV gammas from $^{241}$Am), only alpha events with an energy of $E^{\text{true}}_{\alpha}>5.3\,$MeV were selected. As for the measurements, the error bars represent the standard deviations of the distributions. The reduction of~$\braket{E^{\text{obs}}_{\alpha}}$ and $\braket{\text{DCR}_{\text{r}}}$ is predicted to be stronger for a higher absolute amount of surface charges, particularly at small~$r$, see~\textbf{Fig.}~\ref{graph:galatea_meas_simu_comparison}. A direct comparison of the predicted and measured DCR~values is not meaningful, since they depend on the trace length which is different for measurement and simulation. Therefore, the simulated~$\braket{\text{DCR}_{\text{r}}}$ were scaled with a constant factor which was chosen such that the absolute values roughly match. The predicted radial dependencies describe the measurement results qualitatively well. The predicted dependency of~$\braket{\text{DCR}_{\text{r}}}$ on $r$ describes the measurements also quantitatively for a moderate surface charge of $\sigma=-0.3\cdot10^{10}\,\text{e}/\text{cm}^2$. In contrast, the predicted $\braket{\text{DCR}_{\text{r}}}$ slightly deviate from the measured rates, particularly for $r\gtrsim10\,\text{mm}$. This might be due to the simplicity of the applied DCR model, cf.~\textbf{Eq.}~(\ref{eq:simu_dcr_convolution}).
%\FloatBarrier
\noindent
\FloatBarrier
\noindent

%#############################################################################
\subsubsection{Surface beta events}
\begin{figure}[!h]
\begin{center}
%\hspace{-2cm}
\includegraphics[angle=0,width=0.73\textwidth]{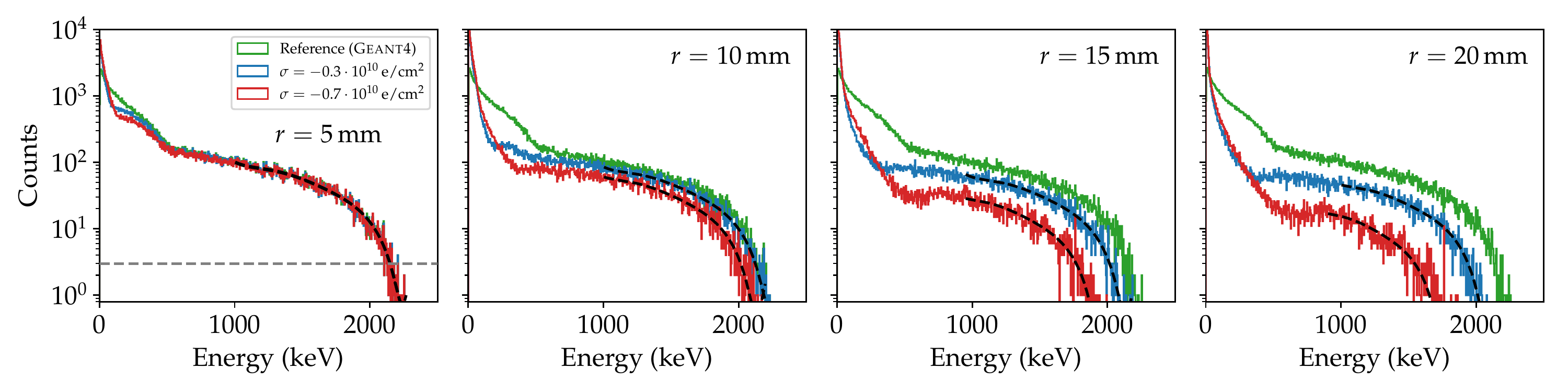}
\caption{Distribution of $E^{\text{true}}_{\beta}$ (\textsc{Geant4} reference, green curve) and $E^{\text{obs}}_{\beta}$ (blue and red curves) for the simulated $^{90}$Sr events at selected radial positions. The~$E^{\text{obs}}_{\beta}$ energy spectra are shown for the signal propagation in the presence of negative surface charges of $\sigma=-0.3, -0.7\cdot10^{10}\,\text{e}/\text{cm}^2$. The spectra were fit with a seventh-order polynomial (black dashed lines) and the intersection with a fixed value was determined.}
\label{graph:galatea_simu_energy_histos_neg_sr90}
\end{center}
\end{figure}

\begin{figure}[!h]
%\myfloatalign
\subfloat[Integral count rate.] {\hspace{-0.2cm}\label{graph:galatea_simu_sr90_rate_evolution_neg}
\includegraphics[width=.5\linewidth]{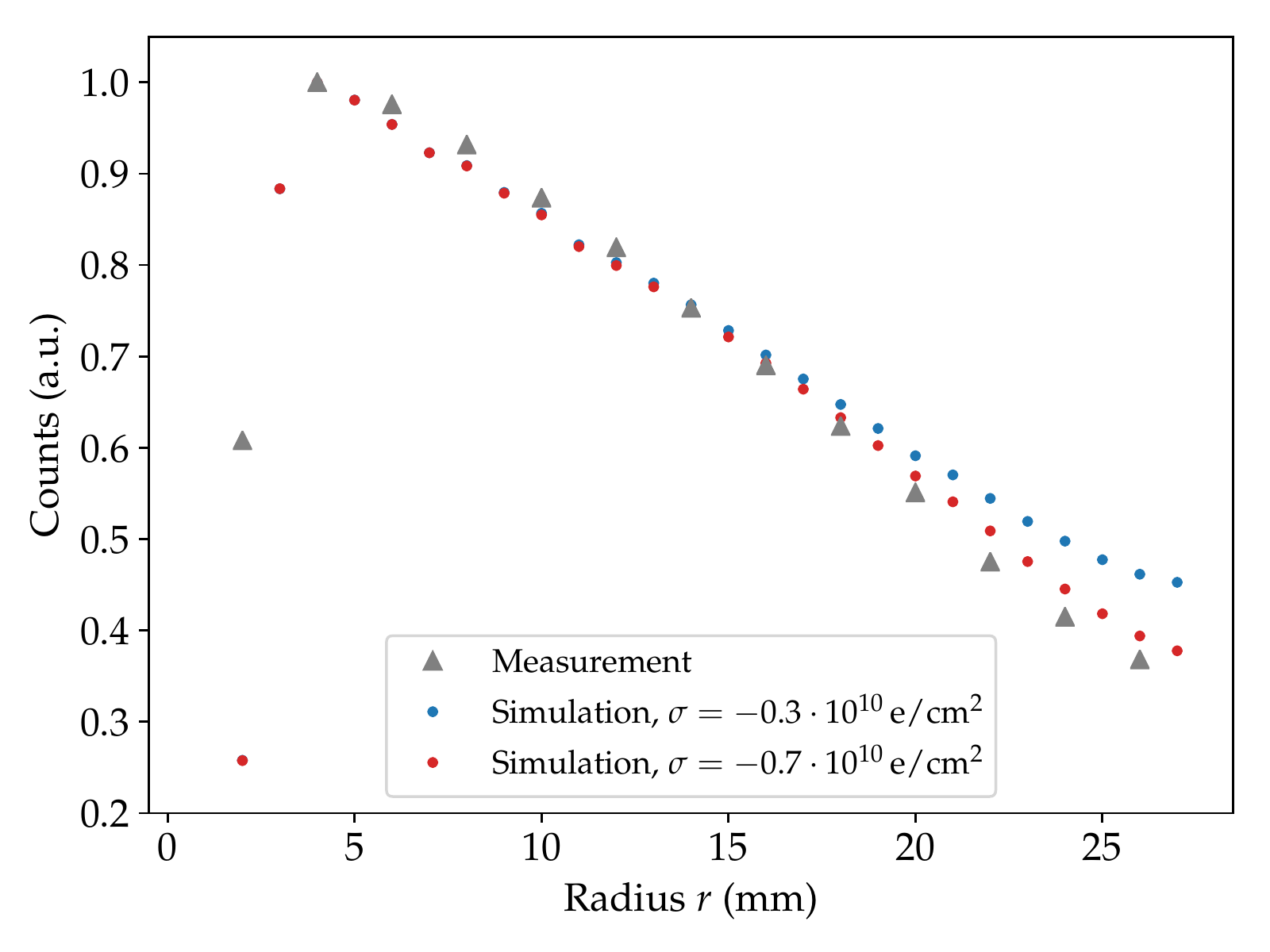}}
\subfloat[Relative endpoints.] {\label{graph:galatea_simu_sr90_rel_endpoints_neg} 
\includegraphics[width=.5\linewidth]{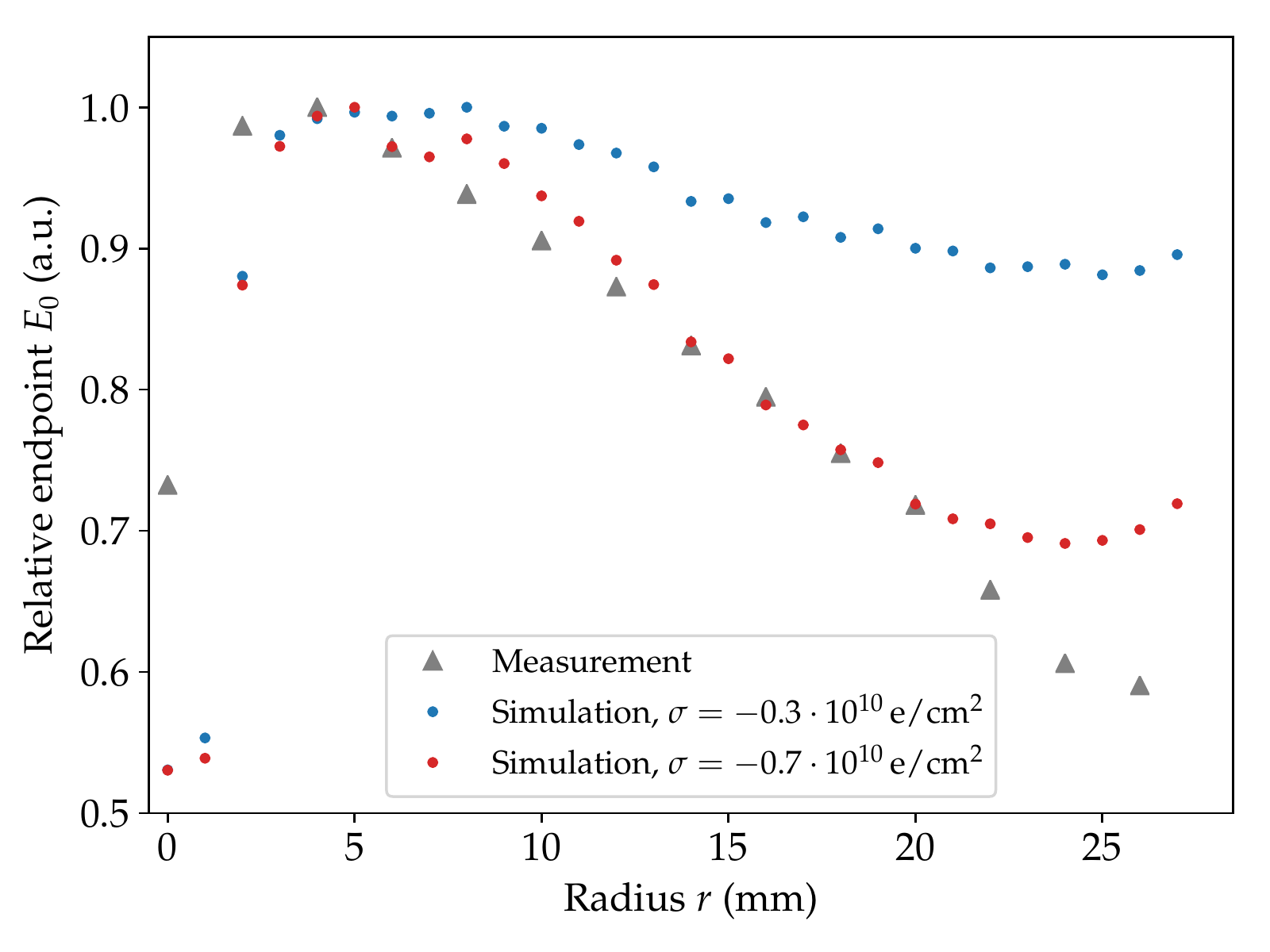}}
\caption{Comparison of measurement to simulation: (a)~integral count rate and (b)~relative endpoints of the $^{90}$Sr $E^{\text{obs}}_{\beta}$ energy spectra. Simulation results are shown for the surface charges $\sigma=-0.3, -0.7\cdot10^{10}\,\text{e}/\text{cm}^2$.}
\label{graph:galatea_simu_sr90_rates_endpoints_neg}
\end{figure}

\begin{figure}[!h]
\begin{center}
%\hspace{-2cm}
\includegraphics[angle=0,width=0.73\textwidth]{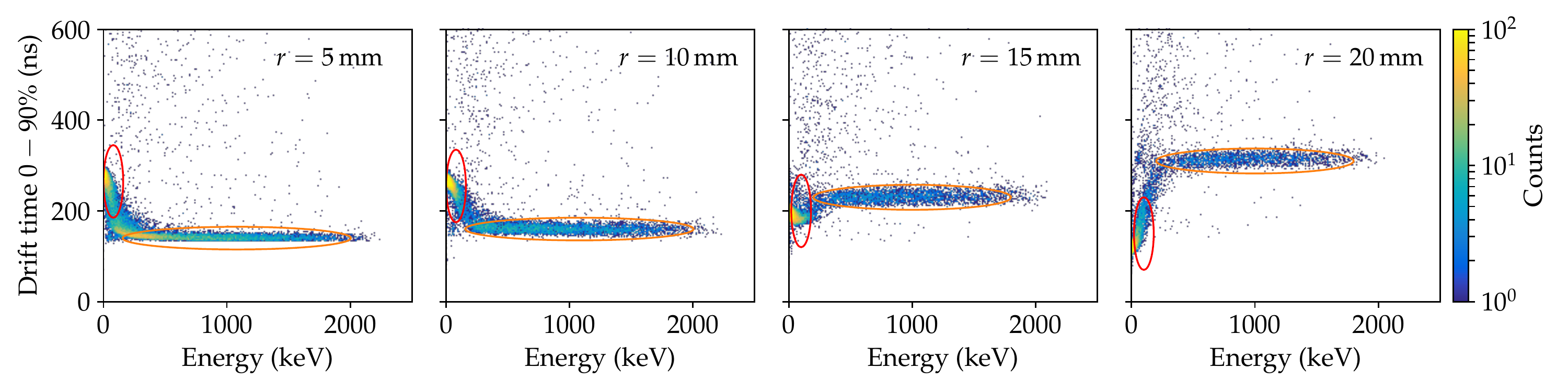}
\caption{Predicted correlation between the drift time and energy~$E^{\text{obs}}_{\beta}$ at selected~$r$ for the simulated $^{90}$Sr events. Simulation results are shown for the surface charge $\sigma=-0.3\cdot10^{10}\,\text{e}/\text{cm}^2$. The red vertical and orange horizontal ellipses in the distributions indicate two event populations.}
\label{graph:galatea_simu_energy_two_populations}
\end{center}
\end{figure}

The analysis of the simulated surface beta events was done in analogy to the analysis of the measurements. The simulated energy spectra in the presence of negative surface charges ($\sigma=-0.3, -0.7\cdot10^{10}\,\text{e}/\text{cm}^2$) are shown in \textbf{Fig.}~\ref{graph:galatea_simu_energy_histos_neg_sr90}. The predicted energy $E^{\text{obs}}_{\beta}$ degrades with increasing~$r$. This is in qualitative agreement with the measurements, cf.~\textbf{Ch.}~\ref{ch:sr90_radial_dependence}. For higher absolute surface charges, the reduction is stronger.
\FloatBarrier
\noindent
The dependence of the integral count rate of~$E^{\text{obs}}_{\beta}$ on~$r$ is shown in \textbf{Fig.}~\ref{graph:galatea_simu_sr90_rate_evolution_neg}. The energy threshold used in the simulation had to be adjusted such that the count rates of the simulation roughly match the measured rates. This might have been necessary because of the simplified drift model which did not take diffusion and self-repulsion into account. Thus, only qualitative statements can be made. The simulation describes the increase of the count rate at small $r$ due to the partial shadowing of the beam spot by the PTFE~bar. Moreover, the simulated integral count rate decreases with~$r$ for $r>5$\,mm. However, it does not drop as steeply as the measured rate. The reduction of~$E^{\text{obs}}_{\beta}$ was also quantified in terms of the shift of the endpoints of the $^{90}$Sr spectra, see \textbf{Fig.}~\ref{graph:galatea_simu_sr90_rel_endpoints_neg}. The spectra were fit with a seventh-order polynomial and the endpoint was approximated as for the measured spectra, cf.~\textbf{Ch.}~\ref{ch:sr90_radial_dependence}. While the simulation describes the measurement reasonably well at small~$r$, it cannot describe the measurement at large~$r$. A better agreement between simulation and measurement is achieved for the higher value of the assumed negative surface charge. This is in contrast to the reasonable description of the alpha events when assuming a lower surface charge.
\FloatBarrier
\noindent
Two event populations were observed in the radial $^{90}$Sr measurements, cf.~\textbf{Ch.}~\ref{ch:sr90_radial_dependence}. The first population consists of events with low energies~$E^{\text{obs}}_{\beta}$, for which the drift time decreases with increasing~$r$. In contrast, the second population consists of events with higher energies~$E^{\text{obs}}_{\beta}$, whose drift time increases with increasing~$r$, see \textbf{Fig.}~\ref{graph:galatea_sr90_before_cuts}. This is validated by the $^{90}$Sr simulations as shown in \textbf{Fig.}~\ref{graph:galatea_simu_energy_two_populations} and can be explained as follows: Events with low $E^{\text{obs}}_{\beta}$ have a small penetration depth. As discussed in \textbf{Ch.}~\ref{ch:charge_efficiency_maps}, they are affected by surface charges and their energy~$E^{\text{obs}}_{\beta}$ is strongly reduced. In contrast, events with a high  $E^{\text{obs}}_{\beta}$ generally have a higher penetration depth. These events are less sensitive to surface effects, i.e.~most of the holes are collected on the p$^+$~contact.
%\FloatBarrier
\noindent

%#############################################################################
\subsection{Discussion and outlook}
The simulations carried out in the scope of this work were based on the drift models for single holes and electrons. The charges were treated as independent and the final signals were calculated as superpositions of the waveforms as expected for isolated point charges. In addition, the environment of the detector was not taken into account. The resulting simulation results were able to describe the data qualitatively well if certain surface charges were assumed. However, some quantitative differences emerged between predictions and data, and the value of the surface charge, which as was to be expected was not always the same. Two effects not taken into account in the simulations could influence the results significantly and will be discussed in the following sections.

\subsubsection{Impact of diffusion and self-repulsion}\label{ch:simu_diffusion_repulsion}
Thermal diffusion and Coulomb self-repulsion are two effects between the charge carriers which are expected to have a significant impact. Since these processes lead to an increase of the size of the charge cloud during its evolution, they could directly influence the fraction of charge carriers
affected by surface effects. 
\\\\
For interactions close to the (passivated) detector surface, the transverse diffusion component and the self-repulsion are of particular importance. The transverse size of the charge cloud initially deposited increases very quickly. The effect becomes stronger for larger and denser energy depositions. In contrast, the longitudinal diffusion is less important because the longitudinal diffusivity is low in a high electric field. Thus, assuming that the field close to the surface is high enough, the charge cloud is expected to become a disk. If the expanding disk drifts parallel to the surface it can eventually intersect with the layer affected by surface effects, even if the original charge cloud did not. This will result in a "smear" of charges which are trapped close to the surface. The lower part of the charge cloud continues drifting, some part can move away and will be collected, some part might be trapped. This results in a modified charge collection behavior compared to the case of independently drifting point charges.
\\\\
First attempts have been made to include these effects in pulse shape simulations. However, this is very challenging. The three-dimensional charge density distributions for both electrons and holes need to be evolved simultaneously. At each time step, a self-consistent electric field has to be recalculated. Moreover, a fine computational grid ($\mathcal{O}(20\,\text{\textmu m})$) and short simulation time steps ($\mathcal{O}(0.2\,\text{ns})$) are required. From the computational point of view, these requirements are very challenging. However, work is ongoing to approximate the effects in two-dimensional calculations as well as to speed up three-dimensional calculations to the point that these effects can be included.

\subsubsection{Impact of the environment}\label{ch:simu_potentials_environment}
\begin{figure}[!h]
\begin{center}
\includegraphics[angle=0,width=0.73\textwidth]{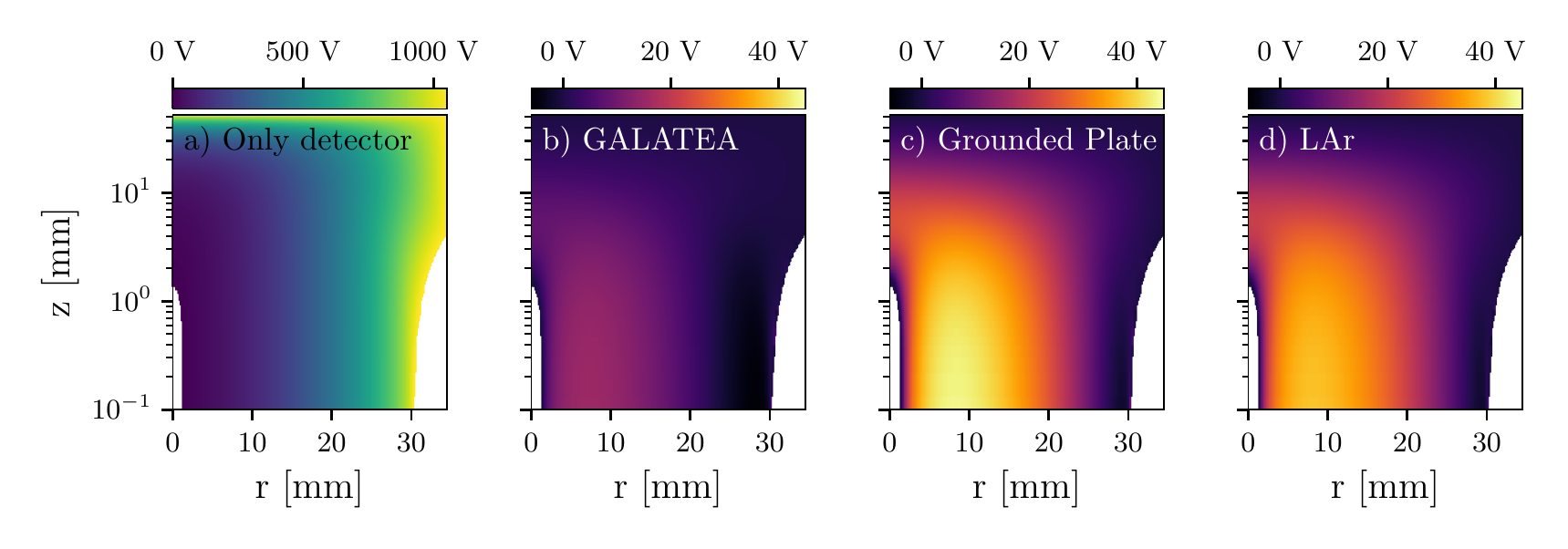}
\caption{Electric potential of the PPC detector under study as calculated with \texttt{SolidStateDetectors.jl}~\cite{SolidStateDetectorsJL2021}: (a)~Potential $\Phi^{\text{bare}}$ as calculated for the bare detector,  (b)-(d) changes in the potential~$\Phi^{\text{env}}-\Phi^{\text{bare}}$ for the detector (b)~mounted in GALATEA, (c)~GALATEA plus a grounded plate above the passivated surface at a distance of~2\,mm, and (d)~submersed in LAr.}
\label{graph:influence_environment}
\end{center}
\end{figure}

The assumed charges on the passivated surface are not the only mechanism to change the potential and thus the field close to this surface. The environment of a detector also has an effect 
on the electric potential. This was investigated for the PPC~detector under study using the newly developed software package \texttt{SolidStateDetectors.jl}~\cite{SolidStateDetectorsJL2021}. The package can calculate the electric potential in and around the detector taking the detector environment into account. The potential was calculated for selected configurations:
\begin{itemize} 
\item[(a)] bare detector, i.e.~the detector surroundings were neglected, reflecting boundary conditions at the surface at~$z=0\,$mm were assumed; 
\item[(b)] detector mounted inside the grounded infrared shield and the detector holding structure of GALATEA;
\item[(c)] detector mounted in GALATEA plus an additional grounded plate above the passivated detector surface at a distance of~2\,mm;
\item[(d)] detector submerged in LAr.
\end{itemize}
The potential~$\Phi^{\text{bare}}$ of the bare detector is shown in~\textbf{Fig.}~\ref{graph:influence_environment}~(a), whereas the changes~$\Phi^{\text{env}}-\Phi^{\text{bare}}$ caused by the different configurations are depicted in panels~(b), (c), and~(d). The environment modifies the electric potential considerably, particularly in the region around the point contact. The grounded plate close to the passivated surface has the largest effect. It amounts to a few percent. The influence of a submersion in liquid argon is almost as strong. Such effects are on the same order of magnitude as the effects calculated for assumed moderate charges on the passivated surface. Consequently, future simulation studies to investigate detector surface effects should take the influence of the surrounding materials on the electric field into account. There is also experimental evidence that the detector under study behaved differently in a different environment~\cite{gruszko2017,arnquist2020}.

%#############################################################################
%#############################################################################
%#############################################################################
\section{Summary and conclusions}\label{ch:conclusions}
The response of a p-type point contact~(PPC) germanium detector to alpha and beta particles was studied in detail to better understand background events as occurring in experiments searching for neutrinoless double beta decay. The results of the measurements in the vacuum test facility GALATEA demonstrate that the structure of events on the passivated detector surface can be explained by effects like surface charges. For both alpha and beta surface events, a radius-dependent reduction of the energy was observed which can be explained by assuming the presence of a negative surface charge. It was also observed that surface alpha events exhibit a delayed charge recovery~(DCR) effect, which can be exploited to effectively reject such events. In dedicated characterization measurements with beta particles, two event populations could be identified. One population could be associated with events having small penetration depths - they are affected by surface effects, whereas the waveforms of the other population with higher penetration depths were found to have no special features. No pronounced DCR~effect was observed for surface beta events. Thus, the identification of beta events on the passivated surface is, if at all, only for part of the events possible. 
\\\\
An extensive simulation campaign was carried out to better understand the results of the surface characterization measurements. Pulse shape parameter maps provided insights into the impact of surface charges on quantities such as the event energy, drift time, etc. In addition, the maps revealed that the influence of a positive surface charge is much less pronounced than that of a negative surface charge. Monte Carlo event simulations in combination with pulse shape simulations were capable of reproducing the measurements well qualitatively. In particular, the simulations confirmed that the measurement can be explained by the presence of a negative surface charge.
\\\\
The presented measurements and corresponding simulations led to a significantly better understanding of PPC~detector surface effects. This serves as a basis to better identify surface events as backgrounds to rare event searches with germanium detectors.

%%%%%%%%%%%%%%%%%%%%%%%%%%%%%%%%%%%%%%%%%%
\vspace{6pt} 

\funding{This work was supported by the Max Planck Society and the Technical University of Munich. F.~Edzards gratefully acknowledges support by the German Academic Scholarship Foundation (Studienstiftung des deutschen Volkes), and S.~Mertens gratefully acknowledges support by the MPRG at TUM program.
\\\\
This material is based upon work supported by the U.S.~Department of Energy, Office of Science, Office of Nuclear Physics under Award Numbers DE-FG02-97ER41041, DE-FG02-97ER41033, DE-AC05-00OR22725, as well as Federal Prime Agreement DE-AC02-05CH11231. This material is based upon work supported by the National Science Foundation under Grant No.~NSF OISE 1743790 and 1812409.}

\acknowledgments{The authors would like to thank R.~J.~Cooper and A.~Poon of Lawrence Berkeley National Laboratory for useful discussions throughout the project, and the \textsc{Majorana} collaboration for the loan of the germanium detector.}

\conflictsofinterest{The authors declare no conflict of interest.} 

\end{paracol}
%%%%%%%%%%%%%%%%%%%%%%%%%%%%%%%%%%%%%%%%%%
% To add notes in main text, please use \endnote{} and un-comment the codes below.
%\begin{adjustwidth}{-5.0cm}{0cm}
%\printendnotes[custom]
%\end{adjustwidth}
%%%%%%%%%%%%%%%%%%%%%%%%%%%%%%%%%%%%%%%%%%
\reftitle{References}


%\externalbibliography{yes}
%\bibliography{your_external_BibTeX_file}

%=====================================
% References, variant B: internal bibliography
%=====================================

%%%%%%%%%%%%%%%%%%%%%%%%%%%%%%%%%%%%%%%%%%
%% for journal Sci
%\reviewreports{\\
%Reviewer 1 comments and authors’ response\\
%Reviewer 2 comments and authors’ response\\
%Reviewer 3 comments and authors’ response
%}
%%%%%%%%%%%%%%%%%%%%%%%%%%%%%%%%%%%%%%%%%%
\end{document}